\documentclass[a4paper,twocolumn,11pt,accepted=2021-05-26]{quantumarticle}
\pdfoutput=1
\usepackage[utf8]{inputenc}
\usepackage[english]{babel}
\usepackage[T1]{fontenc}
\usepackage{amsmath}
\usepackage{hyperref}

\usepackage{tikz}
\usepackage{lipsum}
\usepackage{soul}
\usepackage{float}
\usepackage[sorting=none]{biblatex}
\DeclareBibliographyCategory{needsurl}
\newcommand{\entryneedsurl}[1]{\addtocategory{needsurl}{#1}}
\renewbibmacro*{url+urldate}{%
  \ifcategory{needsurl}
    {\printfield{url}%
     \iffieldundef{urlyear}
       {}
       {\setunit*{\addspace}%
        \printurldate}}
    {}}
\bibliography{database}

\newcommand{\overbar}[1]{\mkern 1.5mu\overline{\mkern-1.5mu#1\mkern-1.5mu}\mkern 1.5mu}

\DeclareMathOperator{\sinc}{sinc}

\begin{document}

\title{Spectrally multimode integrated SU(1,1) interferometer}

\author{A.~Ferreri}
\affiliation{Department of Physics, Paderborn University,
Warburger Strasse 100, D-33098 Paderborn, Germany}
\author{M.~Santandrea}
\affiliation{Department of Physics, Paderborn University,
	Warburger Strasse 100, D-33098 Paderborn, Germany}
\author{M.~Stefszky}
\affiliation{Department of Physics, Paderborn University,
	Warburger Strasse 100, D-33098 Paderborn, Germany}
\author{K.~H.~Luo}
\affiliation{Department of Physics, Paderborn University,
	Warburger Strasse 100, D-33098 Paderborn, Germany}
\author{H.~Herrmann}
\affiliation{Department of Physics, Paderborn University,
	Warburger Strasse 100, D-33098 Paderborn, Germany}
\author{C.~Silberhorn}
\affiliation{Department of Physics, Paderborn University,
Warburger Strasse 100, D-33098 Paderborn, Germany}
\author{P.~R.~Sharapova}
\affiliation{Department of Physics, Paderborn University,
Warburger Strasse 100, D-33098 Paderborn, Germany}

\begin{abstract}
Nonlinear SU(1,1) interferometers are fruitful and promising tools for spectral engineering and precise measurements with phase sensitivity below the classical bound. Such interferometers have been successfully realized in bulk and fiber-based configurations.  However, rapidly developing integrated technologies provide higher efficiencies, smaller footprints, and pave the way to quantum-enhanced on-chip interferometry. In this work, we theoretically realised an integrated architecture of the multimode SU(1,1) interferometer which can be applied to various integrated platforms. The presented interferometer includes a polarization converter between two photon sources and utilizes a continuous-wave (CW) pump. Based on the potassium titanyl phosphate (KTP) platform, we show that this configuration results in almost perfect destructive interference at the output and supersensitivity regions below the classical limit. In addition, we discuss the fundamental difference between single-mode and highly multimode SU(1,1) interferometers in the properties of phase sensitivity and its limits. Finally, we explore how to improve the phase sensitivity by filtering the output radiation and using different seeding states in different modes with various detection strategies.
\end{abstract}
\maketitle

\section{Introduction}
One of the main tasks in Quantum Metrology research is to improve both methods and techniques to estimate the phase sensitivity of interferometers \cite{Chekhova:16}. It is well known that optical phase sensing is generally limited by the noise due to both photons statistics and the quantum nature of light \cite{PhysRevD.23.1693, PhysRevD.26.1817}. The first bound, commonly called shot noise limit (SNL),  describes the noise performance of an ideal classical field and is limited by classical correlations. The SNL is proportional to $1/\sqrt N$, where N is the number of photons in the interferometer \cite{DEMKOWICZDOBRZANSKI2015345, slussarenko2017unconditional}. The second bound condition, known as Heisenberg limit (HL), stems from the quantum uncertainty principle and is proportional to $1/N$ \cite{PhysRevA.55.2598}.

The shot noise limit can be overcome by using quantum states in a typical Mach-Zehnder interferometer. For example, one can insert a Fock state  or the more exotic NOON state \cite{doi:10.1080/00107510802091298} into input channels of such an interferometer. However, the preparation of such exotic states is a challenging task. The nonlinear SU(1,1) interferometers can beat the SNL even without using quantum states as inputs \cite{PhysRevA.33.4033}. This type of interferometer consists of two nonlinear processes \cite{Seyfarth2020wignerfunction} (sometimes one in a truncated version \cite{Gupta:18, https://doi.org/10.1002/qute.201900138}), such as four-wave-mixing (FWM) or parametric down-conversion (PDC).

The phase sensitivity of this class of interferometers has been investigated for spectrally single-mode sources \cite{manceau2017improving,PhysRevApplied.10.064046, Xiao_2019}, including gain-unbalanced \cite{PhysRevLett.119.223604, PhysRevA.96.053863} and fully quantum three-mode  \cite{Fl_rez_2018} configurations, as well as configurations with different input states, such as coherent light \cite{Plick_2010}, squeezed states \cite{PhysRevA.95.063843}, a mixture of coherent and squeezing states \cite{Li_2014, Adhikari:18,Guo:18} and a mixture of thermal and squeezed states \cite{Ma:18}. These studies have shown that it is generally possible to overcome the shot noise limit and even reach the Heisenberg limit.
Moreover, SU(1,1) interferometers can provide different benefits with respect to other interferometers, not only in terms of high-precision measurements, but also because of the possibility to perform a joint measurement of multiple observables \cite{PhysRevA.97.052127} and due to the advantages of robustness against external losses, namely losses due to an inefficient detection system \cite{Li:18, PhysRevA.86.023844, doi:10.1063/1.4960585,PhysRevA.98.023803}.

Nevertheless, the single-mode picture, commonly used to analyse SU(1,1) interferometers, is merely an approximation of what can realistically be obtained in experiments. Indeed,  PDC sources generate photon pairs with finite spectral and spatial bandwidths and  generally in more than a single mode. Therefore, the complete description of SU(1,1) interferometers should account for the presence of such spatial \cite{PhysRevA.91.043816, Frascella_2019,  PhysRevA.101.053843} and spectral modes \cite{PhysRevA.97.053827, PhysRevLett.117.183601, PhysRevX.10.031063} in order to correctly engineer the system \cite{paterova2020nonlinear} and accordingly maximize the phase sensitivity \cite{Frascella:19}.


The theoretical analysis of SU(1,1) interferometers has been limited so far to free space bulk \cite{doi:10.1063/5.0004873} or fiber-based systems \cite{Su:19, doi:10.1063/1.5048198}. However, it is of particular interest to extend such analysis to other integrated systems \cite{obrien2013, tanzilli2012, Ono:19}, which can provide higher efficiencies, smaller footprints and pave the way to quantum-enhanced on-chip interferometry. Such integrated systems have been already utilized for instance in boson sampling \cite{Spring798} and other elements of quantum computing, such as quantum walk \cite{Peruzzo1500} and Hong-Ou-Mandel interference \cite{Sharapova_2017}.

In this work, we present a theoretical description of the multimode integrated SU(1,1) interferometer. The presented architecture can be realized on various integrated platforms. As an example, we investigate the spectral properties of the monolithic SU(1,1) interferometer based on the  potassium titanyl phosphate (KTP) platform and demonstrate conditions for obtaining the phase sensitivity below the shot noise limit.

The paper is organized as follows: in the second section, a brief introduction of the theoretical model and the phase sensitivity calculation strategy are derived and discussed.
In the third section, we outline two different sample designs: the conventional bulk architecture implemented on the integrated platform and a new developed integrated design. We demonstrate that this new design allows the phase sensitivity to overcome the SNL and generate supersensitivity bands. In the fourth section, we show how phase sensitivity can be improved by using  a frequency filter. In the last section, we extend the analysis by considering non-vacuum input states along with different detection strategies, i.e. direct detection and homodyne detection.


\section{Theoretical model}

The Hamiltonian describing the whole interference process inside the SU(1,1) interferometer can be represented via the joint spectral amplitude (JSA) $F(\omega_s,\omega_i)$ \cite{PhysRevA.97.053827}:
\begin{equation}
\hat H=i \hbar \Gamma \int d\omega_s d\omega_i F(\omega_s,\omega_i)\hat a_s^\dagger \hat a_i^\dagger+h.c.,
\label{Ham1}
\end{equation}
where $\Gamma$ is the coupling constant containing the second order susceptibility $\chi^{(2)}$, the pump intensity $I$ and the length of PDC sections; $\omega_s$ and $\omega_i$ are the frequencies of the signal and idler photons, respectively. Here we neglect the effects of time ordering. However, such an approximation is well satisfied for the SU (1,1) interferometer due to the effective narrowing of the spectrum because of the nonlinear interference \cite{PhysRevLett.117.183601}. The joint spectral amplitude depends on both the spectral properties of the pump laser and the geometry of the system, and it is generally expressed as:
\begin{equation}\begin{split}
F(\omega_s,\omega_i)=C\,\alpha(\omega_s,\omega_i)f(\omega_s,\omega_i),
\label{jsa}
\end{split}\end{equation}
where $C$ is the normalization constant. The function $\alpha(\omega_s,\omega_i)$ depends on the spectral properties of the pump and has a Gaussian profile:
\begin{equation}
\alpha(\omega_s,\omega_i)=e^{-\frac{(\omega_s+\omega_i-\omega_p)^2 \tau^2}{2}},
\label{alpha}
\end{equation}
where $ \tau$ is the pulse duration of the pump laser, $\omega_p$ is the pump frequency.

The function $f(\omega_s,\omega_i)$ is the phase matching function. In a general form, for both integrated circuits with two PDC sections of equal length $L$ investigated in this work, the function $f(\omega_s,\omega_i)$ is given by \entryneedsurl{klyshko1993ramsey}\entryneedsurl{klyshko1994parametric}\cite{klyshko1993ramsey, klyshko1994parametric, Santandrea_2019, Helmfrid:93}:
\begin{equation}
f(\omega_s,\omega_i)=\frac{1}{2 L}\int_0^{2L+l}dz \,g(z)\,e^{i\int_0^z d\xi \Delta k(\xi)},
\label{ef}
\end{equation}
where $g(z)$ is  the spatial profile of the second-order nonlinear susceptibility \cite{HUM2007180},  $l$ is the distance between two PDC sections, $\Delta k (\xi)$  is the phase matching profile, corresponding to the type-II PDC process.

 

By solving the Heisenberg equations for the Hamiltonian in Eq.(1), one can calculate the mean number of signal photons $\langle N \rangle$ and its variance $\langle \Delta^2N \rangle$ for a given input state of the interferometer, see Appendix.
Consequently, the phase sensitivity can be calculated as:  
\begin{equation}
\lvert\Delta\phi\rvert=\bigg\lvert\frac{\Delta N}{\partial \langle N \rangle/\partial\phi}\bigg\rvert,
\label{psdef}
\end{equation}
where $\phi$ is the phase shift undergone by photons inside the interferometer. 
Interferometers are characterized by the supersensitivity regions if the phase sensitivity overcomes the shot noise limit:  
\begin{equation}
\Delta\phi_{SNL}=\frac{1}{\sqrt {\langle N_{in} \rangle}},
\label{snl}
\end{equation}
where $N_{in}$ is the number of interfering photons inside the interferometer. Starting from the next section we will use the phase sensitivity normalized in respect to the shot noise limit.

\section{Interferometer optimization}

In contrast to bulk optics, where temporal dynamics of the PDC photons can often be neglected, the long interaction lengths  in integrated SU(1,1) require a careful design that takes into account the temporal walk-off and stretching of pulses in the material due to group velocity dispersion.

\subsection{The conventional bulk architecture implemented on the integrated platform}

\begin{figure}[H]
	\centering
	\includegraphics[width=1\linewidth]{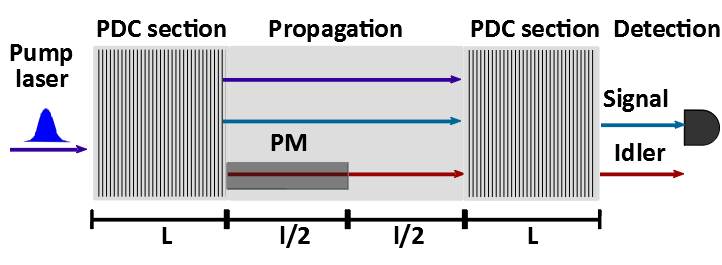}
	\caption{A schematic model of the collinear integrated SU(1,1) interferometer. An incoming pump laser interacts with a periodic poled PDC section, generating a signal-idler photon pairs. In the propagation path, idler photon undergoes a phase modulation (PM). Afterwards, both the generated photons and the pump beam interact in the second PDC section. Finally, the signal photon is detected. Different colours for signal and idler photons are used in order to distinguish vertical and horizontal polarizations. }
	\label{fignc}
\end{figure}

The conventional bulk optics design of the SU(1,1) interferometer, transferred to a fully integrated  platform, is shown in Fig.\ref{fignc}.

\begin{figure*}
	\centering
	\includegraphics[width=1\linewidth]{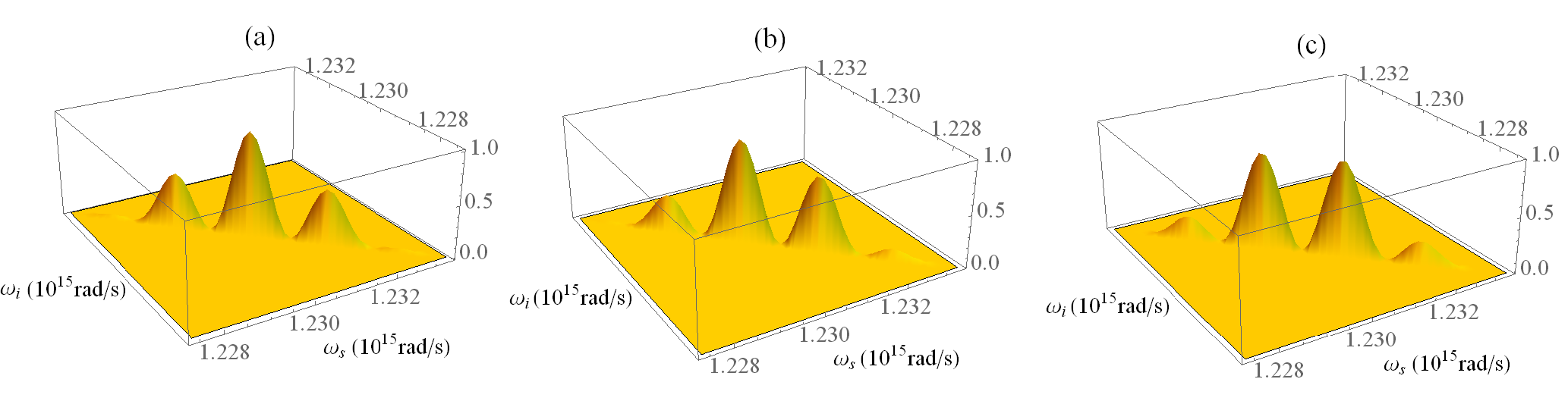}
	\caption{The joint spectral intensity  $\vert F(\omega_s, \omega_i)\vert^2$ depending on the phase implemented by the phase modulator: (a) $\phi=0$, (b) $\phi=\pi/2$ and (c) $\phi=\pi$. The following parameters are chosen: the CW laser, the KTP (potassium titanyl phosphate) platform, the pump wavelength $\lambda_p=766 nm$, the crystal length $L=8mm$, the distance between the two poling sections  $l=10mm$, the period of poling $\Lambda=126\mu m$.}
	\label{anglejsa}
\end{figure*}
In this scheme, a pump laser interacts with the first crystal creating a pair of signal-idler photons.  Along the propagation path, the phase of one or more of the fields can be modulated via an active element. Henceforth, we consider only a phase modulation of the e-polarized idler field, which can be achieved e.g. via electrooptic modulation, that changes the refractive index profile $n(\omega_i)$. For this reason, in the modulator region we will consider a new idler wavevector $k'_i$, such that $\delta k_e = k_e-k'_e$ is the variation of the idler wavevector induced by the phase modulator. Afterwards, the three fields interact in the second PDC source and finally the generated photons are detected. 
The spatial profile $g(z)$ of the second-order nonlinear susceptibility of the sample under consideration can be written as:
\begin{widetext}
\begin{equation}
g(z)=\begin{cases}square_\Lambda(z) & 0<z<L \vee L+l<z<2L+l \\1 & L<z<l+L \end{cases}, 
\label{grate}
\end{equation}
\end{widetext}
where $square_\Lambda(z)$ is the periodic square function that oscillates between -1 and +1 with periodicity $\Lambda$ and describes periodic poling.
Taking into account the first term of the Fourier expansion of  Eq.(\ref{grate}) in the poling region, the JSA of the sample, as calculated by equations (\ref{jsa}), (\ref{ef}), is given by:  
\begin{widetext}
\begin{equation}\begin{split}
F(\omega_s,\omega_i)=\frac{C}{2L}\alpha(\omega_s,\omega_i)\bigg(\int_{0}^{L} dz\, e^{i\frac{2\pi z}{\Lambda}}e^{i\int_{0}^{z}d\xi \Delta k(\xi)}
+\int_{L}^{L+l}dz\, e^{i\int_{0}^{z}d\xi\Delta k(\xi)}+\int_{L+l}^{2L+l}dz\,e^{i\frac{2\pi z}{\Lambda}} e^{i\int_{0}^{z}d\xi\Delta k(\xi)}\bigg)\\
=C\alpha(\omega_s,\omega_i)\sinc\left[\frac{\Delta \beta L}{2}\right] 
\times \cos\left(\frac{2\Delta \beta L+\Delta \beta l+\Delta \beta' l}{4}\right)e^{i(\Delta \beta L+\Delta \beta l/4+\Delta \beta' l/4)},
\label{fnocomp}
\end{split}\end{equation}
\end{widetext}
where $\Delta \beta=k_o(\omega_p)-k_o(\omega_s)-k_e(\omega_i)+2\pi/\Lambda$ is the phase mismatch, compensated by the poling period $\Lambda$,  $\Delta \beta'= k_o(\omega_p) - k_o(\omega_s) - k'_e(\omega_i)+2\pi/\Lambda$ is the phase mismatch in the modulator region,  while indices "o" and "e" denote the "ordinary" and "extraordinary" polarization, respectively. 
The first integral in Eq.(\ref{fnocomp}) contains the phase matching function describing the first PDC section, whereas the third integral describes both the propagation of all beams inside the interferometer and the phase matching in the second PDC section. 
Note, that terms corresponding to the linear propagation in the waveguide are also characterized by the mismatch $\Delta \beta$. This is a consequence of the fact that, in our consideration, the poling profile of the second PDC section is in phase with the poling profile of the first PDC section. This does not apply to two independent  gratings.
The second integral corresponds to the generation of PDC light between the two poling regions, the contribution of this integral is neglected, since the absence of the grid makes the integrand a fast oscillating function resulting in a negligible magnitude of the integral.

The variation of $n(\omega_i)$ can be experimentally realized by applying an external voltage. In theory, since the phase solely affects the idler photon, we can write
\begin{equation}
\frac{\Delta\beta' l}{2}=\frac{\Delta\beta l}{2}+\frac{k_e(\omega_i)l-k_e'(\omega_i)l}{2}=\frac{\Delta\beta l}{2}+\frac{\delta k_e(\omega_i)l}{2}.
\label{deltab}
\end{equation}
The average phase imparted by the phase modulator to the idler field can be then defined as $\phi=\delta k_{e}(\omega_p/2)l/2$, and Eq.(9) can be rewritten as: 
\begin{equation}
\frac{\Delta\beta' l}{2}=\frac{\Delta\beta l}{2}+\frac{\delta k_e(\omega_i)\phi}{\delta k_e(\omega_p/2)}.
\label{deltabb}
\end{equation}

Although the mathematical model presented so far does not depend on the waveguide material explicitly, here onwards we will focus on an integrated SU(1,1) interferometer based on the KTP platform. Fig.\ref{anglejsa} shows how the applied phase shift drastically modifies the shape of the joint spectral intensity (JSI) $\vert F(\omega_s, \omega_i)\vert^2$ and leads to the modulation of both signal and idler spectra. 
 It is important to notice that, in such configuration, for any values of phase $\phi$ a non-vanishing mean number of photons exists: due to different polarizations and, as a consequence, different refractive index profiles (and group velocities) of signal and idler photons, there is no $\phi$ values that can lead to perfect destructive interference in the JSA. Since destructive interference is required for the observation of supersensitivity \cite{Chekhova:16}, the proposed design based on the conventional bulk SU(1,1) architecture is inadequate for photon sensing applications in integrated optics.

Nevertheless, this problem can be overcome as shown in the next section.

\subsection{ The new integrated architecture}

\begin{figure}[H]
	\centering
	\includegraphics[width=1\linewidth]{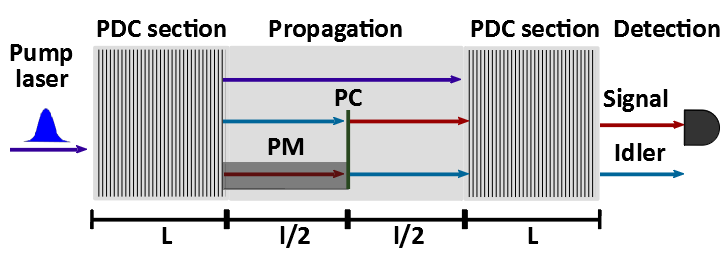}
	\caption{A schematic model of the integrated SU(1,1) interferometer with the time delay compensation. An incoming pump laser interacts with a periodic poled PDC section, where signal-idler photons pairs are generated. The idler photon undergoes a phase modulation (PM). Afterwards, a polarization converter (PC) located in the middle of the device switches polarizations of signal and idler photons. Finally, all beams interact in the second PDC section. In the end, the signal photon is detected. Different colours are used in order to distinguish vertical (red) and horizontal (blue) polarizations.}
	\label{figwc}
\end{figure}

To obtain almost ideal destructive interference at the output of an integrated SU (1,1) interferometer, the compensation for the time delay between signal and idler photons, arising due to the difference in their group velocities, is strictly required.
The integrated SU(1,1) interferometer which takes into account such a compensation is depicted in Fig. \ref{figwc}.  This interferometer has the same layout as the one previously presented, with the addition of a polarisation converter (PC) at the centre of the device. The PC switches polarizations of the signal and the idler photons in order to compensate for the arising time delay.

The JSA characterizing such interferometer can be obtained by following the same procedure as before, but taking into account the change in the phase matching after PC:  
\begin{equation}\begin{split}
F(\omega_s,\omega_i)=\\=\frac{C}{2}\alpha(\omega_s,\omega_i)\bigg[\sinc\bigg[\frac{ \Delta \beta L}{2}\bigg]e^{i\Delta \beta L/2}+\sinc\bigg[\frac{ \overbar{\Delta \beta} L}{2}\bigg]\\
\times\exp\bigg\{i\bigg(\frac{\overbar{\Delta \beta}L}{2} + \Delta \beta L+\frac{(\Delta \beta'+\overbar{\Delta \beta}) l}{2}\bigg)\bigg\}\bigg],
\label{JSAcomp}
\end{split}\end{equation}
where $\overbar{\Delta \beta}=k_o(\omega_p)-k_e(\omega_s)-k_o(\omega_i)+2\pi/\Lambda$ is the phase matching after the polarization switch. Note that the polarization of the signal and idler fields in $\overbar{\Delta \beta}$ is different compared to $\Delta\beta$.
\begin{figure*}
	\centering
	\includegraphics[width=1\linewidth]{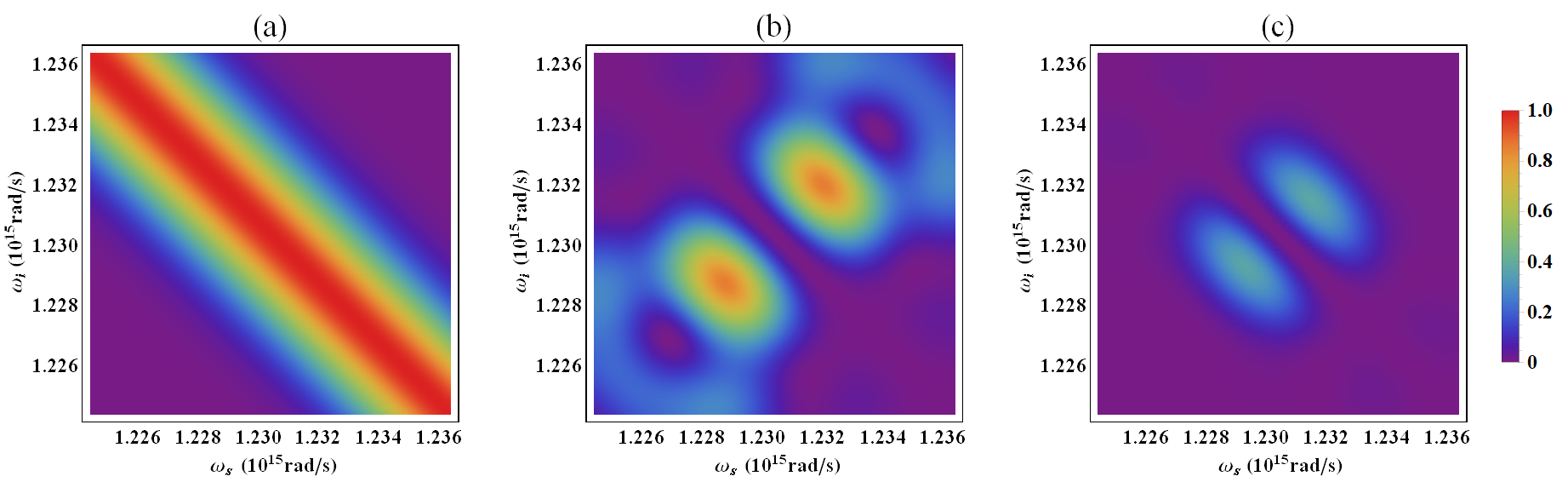}
	\caption{ The KTP platform. (a) The pump spectral bandwidth $\alpha(\omega_s, \omega_i)$,   (b) the phase matching function $\lvert f(\omega_s,\omega_i)\rvert$  and  (c) the joint spectral intensity $\lvert F(\omega_s,\omega_i)\rvert^2$ in the pulsed regime when $\phi\simeq\pi$. The pulse duration is $\tau=0.35ps$, $L=8mm$, $\lambda_p=766 nm$,  $l=10mm$, $\Lambda=126\mu m$. By increasing the pulse duration, $\alpha(\omega_s,\omega_i)$ gets narrower with respect to the $\omega_s= \omega_p -\omega_i$ diagonal, this leads to zero overlap between $\alpha(\omega_s,\omega_i)$ and $f(\omega_s,\omega_i)$ and purely destructive interference.}
	\label{JSAplot}
\end{figure*}


To avoid additional compensation for the time delay between signal/idler and pump photons and, at the same time, to reduce the uncompensated second-order effects, we use a CW laser. Indeed, the stretching of the pump pulse is inversely proportional to the square of the pulse duration. Therefore, in the CW pump case, this stretching can be neglected. The high degree of correlation between the signal and the idler photons determined by the employment of the CW laser allows us to assume $\omega_i=\omega_p-\omega_s$. Substituting this expression into the Taylor expansion of $\Delta \beta$ in the proximity of $\omega_p/2$ and considering terms up to the first order inclusive, we have:
\begin{equation}\begin{split}
\Delta \beta\approx \underbrace{k_o(\omega_p)-k_o(\omega_p/2)-k_e(\omega_p/2)+\frac{2\pi}{\Lambda}}_{=0} \\
-\frac{\partial k_o}{\partial \omega}\mid_{\frac{\omega_p}{2}}(\omega_s-\omega_p/2)-\frac{\partial k_e}{\partial \omega} \mid_{\frac{\omega_p}{2}}(\omega_p/2-\omega_s).
\label{deltak}
\end{split}\end{equation}
Since the first line in Eq.(\ref{deltak})  vanishes due to the definition of periodic poling, the phase matching is simply: 
\begin{equation}
\Delta \beta\approx -\frac{\Omega}{v_o(\omega_p/2)}+\frac{\Omega}{v_e(\omega_p/2)},
\label{do}
\end{equation}
where the frequency detuning $\Omega=\omega_s-\omega_p/2$ and the group velocity $v=(\partial k/\partial \omega)^{-1}$ are introduced. By applying the same strategy to $\overbar{\Delta \beta}$ we finally obtain:
\begin{equation}
\overbar{\Delta \beta}\approx -\frac{\Omega}{v_e(\omega_p/2)}+\frac{\Omega}{v_o(\omega_p/2)},
\end{equation}
which means that $\overbar{\Delta \beta}\simeq -\Delta \beta$.

By substituting this last equality in Eq.(\ref{JSAcomp}), we get:
\begin{widetext}
\begin{equation}\begin{split}
F(\omega_s,\omega_i)\simeq C\,\delta(\omega_p-\omega_s-\omega_i)\sinc\left[\frac{ \Delta \beta L}{2}\right] \cos\left[\frac{\delta k_e(\omega_i)\phi}{2\delta k_e(\omega_p/2)}\right]\exp\left\{\frac{i}{2}\left(\Delta \beta L+\frac{\delta k_e(\omega_i)\phi}{2\delta k_e(\omega_p/2)}\right)\right\},
\label{JSAcomp2}
\end{split}\end{equation}
\end{widetext}
 where $\delta(\omega_p-\omega_s-\omega_i)$ is the Dirac delta function describing the CW pump. Numerically, the CW regime is achieved by setting the pulse duration $\tau$ in Eq.(\ref{JSAcomp}) so long that the pump can cover entirely the whole length of the interferometer; this condition is mathematically expressed as $c\tau\gg 2L+l$. 

Since the relation $\frac{\delta k_e(\omega_i)}{\delta k_e(\omega_p/2)} \simeq 1$ in the vicinity of  $\omega_p/2$, it can be clearly seen that Eq.(\ref{JSAcomp2}) results in the JSA of a single PDC process modulated by $\cos(\phi/2)$. Due to this modulation it is possible to achieve almost perfect destructive interference in a large bandwidth around $\omega_p/2$ in the JSA using specific values of $\phi$ (which is in stark contrast to the non-compensated configuration, whose JSA profiles are presented in Fig.\ref{anglejsa}).


This regime, where the JSA is drastically attenuated, can be achieved with relative simplicity thanks to the spectral feature of our device based on the KTP (potassium titanyl phosphate) platform.
 In this framework, the natural orientation of the phase matching function $f(\omega_s,\omega_i)$ in KTP, whose slope is typically positive in the '$\omega_s-\omega_i$' diagram, together with the presence of a PC make sure that the phase matching function of the  SU(1,1) interferometer lies along the $\omega_s=\omega_i$ diagonal (Fig. \ref{JSAplot}b); moreover, the additional phase induces a modulation leading to a destructive interference at $\phi=\pi$, in which the phase matching function vanishes along the antidiagonal $\omega_p=\omega_s+\omega_i$.

Although the presence of the phase modulator and the PC changes the structure of the phase matching function and inhibits the output light intensity, the pulsed regime with a broad pump spectrum $\alpha(\omega_s,\omega_i)$, see Fig.\ref{JSAplot}a, leads to the generation of a notable number of photons and a non-vanishing JSA with two high peaks, as demonstrated in Fig.\ref{JSAplot}c. However, such residual light can be further inhibited by decreasing the pump spectral bandwidth (using a CW laser), in order to dramatically reduce the overlapping region between $\alpha(\omega_s,\omega_i)$ and $f(\omega_s,\omega_i)$ and minimize the number of generated photons.  In this context, the use of the CW pump laser is an important and necessary requirement in order to achieve a notable reduction of the mean number of photons.

\begin{figure}[H]
	\centering
	\includegraphics[width=1\linewidth]{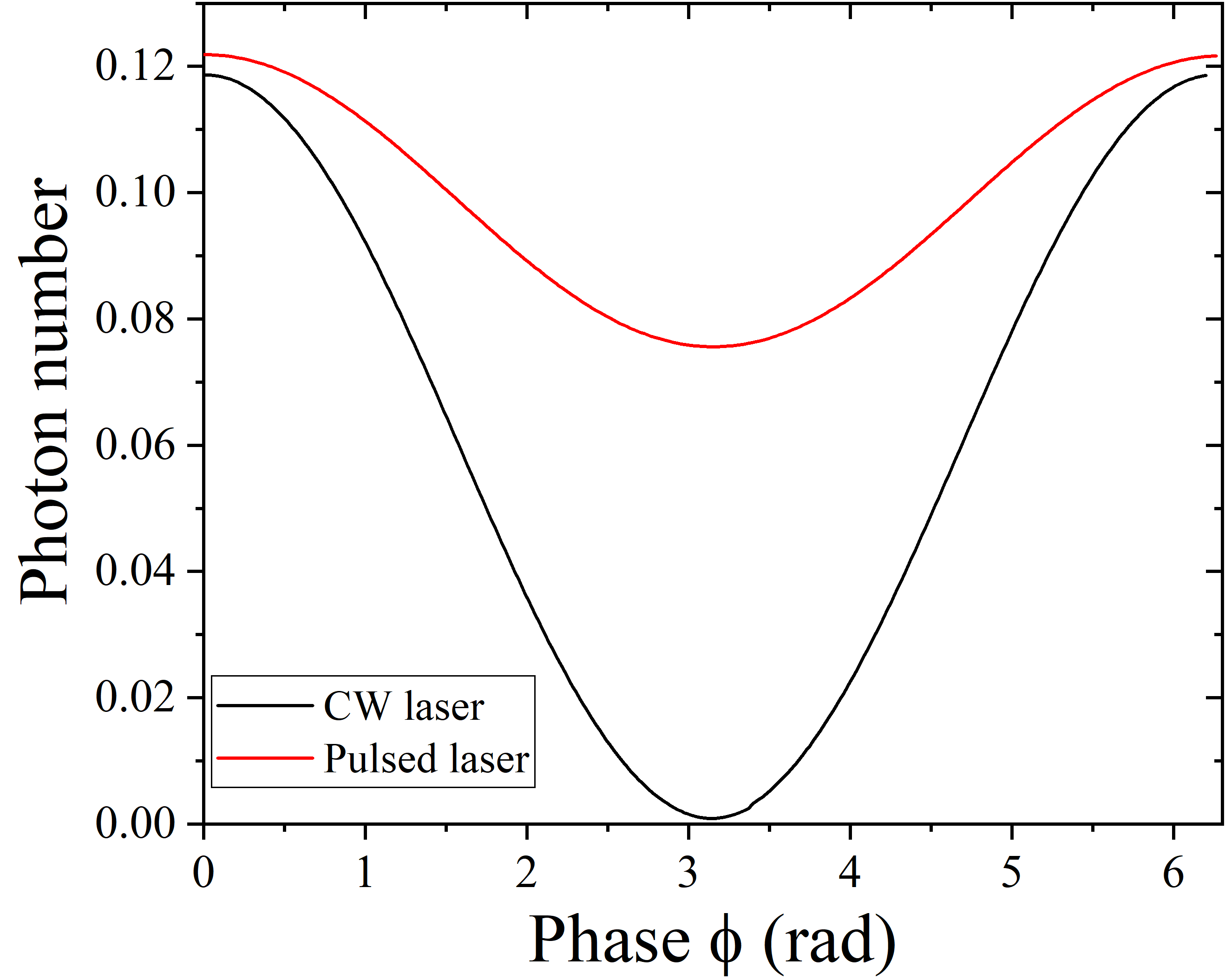}
	\caption{The comparison between the number of photons in the case of pulsed laser (the pulse duration $\tau=0.35 ps$, red line) and CW regime. In order to have a situation with the resulting number of photons are nearly identical at $\phi=0$, different pump intensities are used for the two pumping regimes. The following parameters are chosen: $L=8mm$,  $\lambda_p=766 nm$, $l=10mm$, $\Lambda=126\mu m$.}
	\label{Nsinglevsmulti}
\end{figure}

The interference profile, namely the integral number of photons depending on the phase, is presented in Fig.\ref{Nsinglevsmulti} for two different types of pump lasers: the pulsed and the CW laser. A comparison of the two curves was performed by making use of different pump powers in order to achieve a similar number of output photons.  One can clearly observe that the use of the CW laser drastically increases the visibility of the modulation and allows the photon number to drop close to zero.
The behaviour of the phase sensitivity, defined in Eq.(\ref{psdef}) and normalized to the shot noise limit Eq.(\ref{snl}), is presented in Fig.\ref{noseeding} for different gains, where we introduce the gain parameter $\gamma=G(0)\sqrt{\lambda_1(0)}$, see Appendix.
In this framework, the variation of $\gamma$  is determined by changing the coupling constant $\Gamma$. This allows to span a range of mean photon number between $\langle N(\phi=0)\rangle\approx0.12$ for $\gamma\simeq0.04$, and $\langle N(\phi=0)\rangle\approx10^9$ for $\gamma\simeq10.0$.

\begin{figure}[H]
	\centering
	\includegraphics[width=1\linewidth]{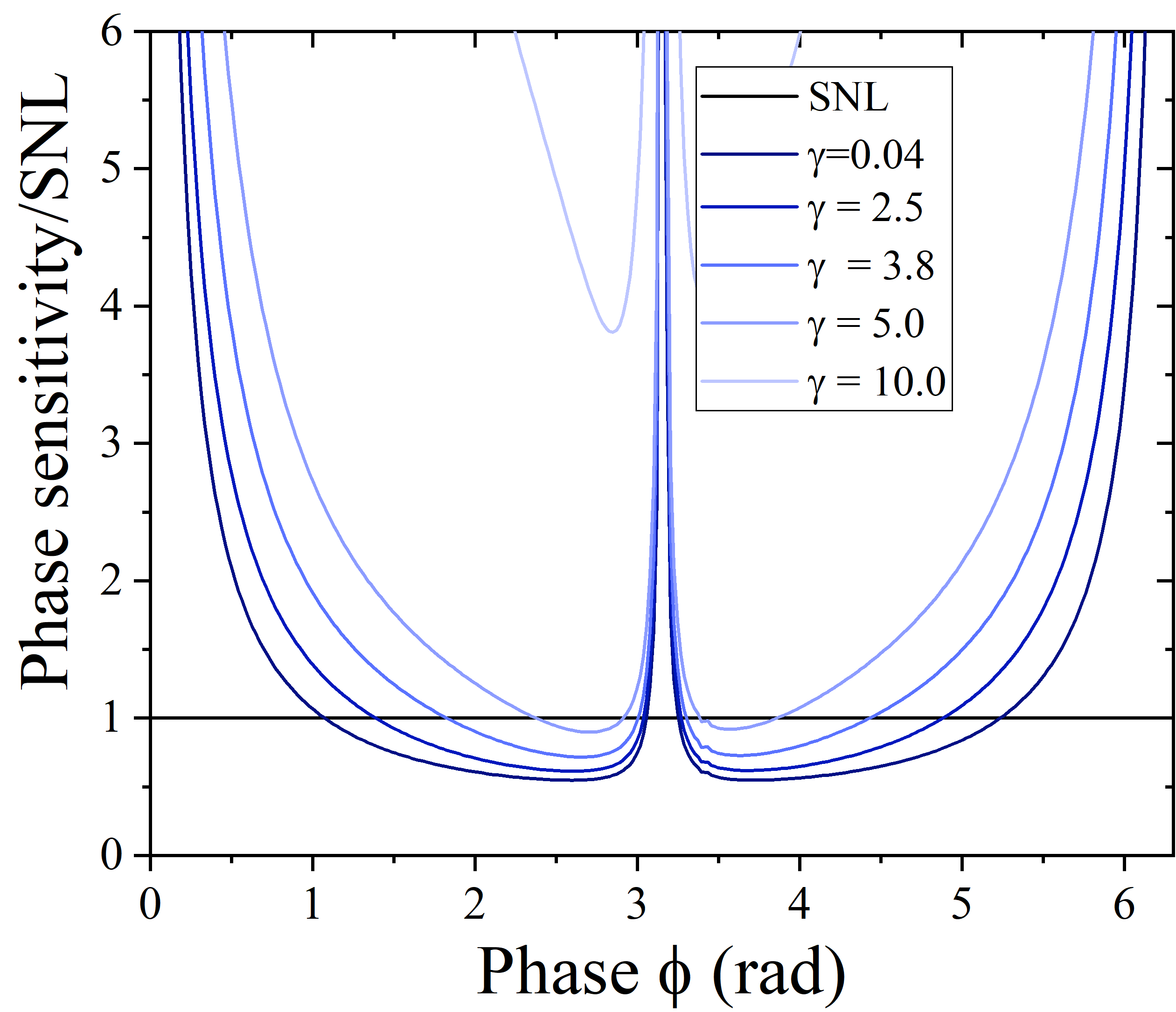}
	\caption{The phase sensitivity normalized to SNL versus phase at different gains. The SNL is shown by the black line. The following parameters are chosen and fixed for all further calculations: the CW laser,  $L=8mm$,  $\lambda_p=766 nm$, $l=10mm$, $\Lambda=126\mu m$.}
	\label{noseeding}
\end{figure}
\begin{figure}[H]
	\centering
	\includegraphics[width=1\linewidth]{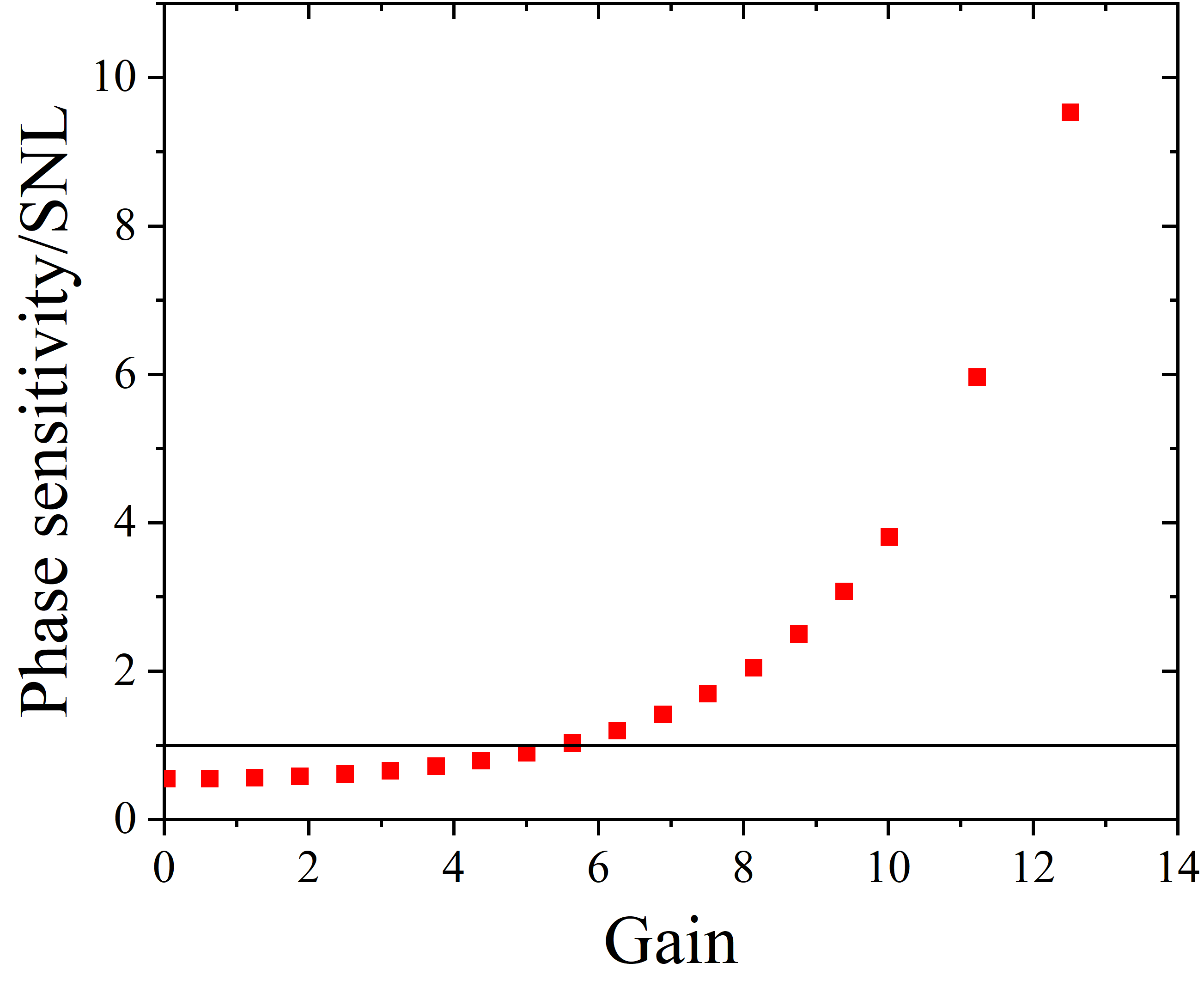}
	\caption{The minimum values of the normalized phase sensitivity presented in Fig.\ref{noseeding} versus gain $\gamma$. The higher the gain, the faster the phase sensitivity grows. The SNL is plotted in black.}
	\label{psvsgain}
\end{figure}
The central peak at $\phi=\pi$  in Fig.\ref{noseeding} is a feature of the multimode structure of the considered interferometer and stems from the non-perfect interference process: due to the fact that the relation $\overbar{\Delta \beta}\simeq-\Delta \beta$ holds only for the first order of the Taylor expansion around the central frequency $\omega_p/2$, and not strictly for large frequency detuning $\lvert\Omega\rvert>0$. As a result, the mean number of photons and its variance are not  entirely  zero for this value of $\phi$, whereas the derivative $\partial \langle N \rangle/\partial\phi$ is exactly zero. 
This leads to a degradation in phase sensitivity with increasing gain and reveals one of the deepest differences between phase sensitivity analysis for single-mode plane-wave interferometers, typically presented in the literature
\cite{manceau2017improving,Plick_2010,PhysRevA.95.063843, Li_2014}, and the multimode scenario considered here, where the broad spectral distribution of the generated light is taken into account. 
Although the proposed scheme improves the visibility of interference pattern significantly, a perfect interference can never occur because of the complex material dispersion of the integrated device.

\begin{figure*}
	\centering
	\includegraphics[width=0.8\linewidth]{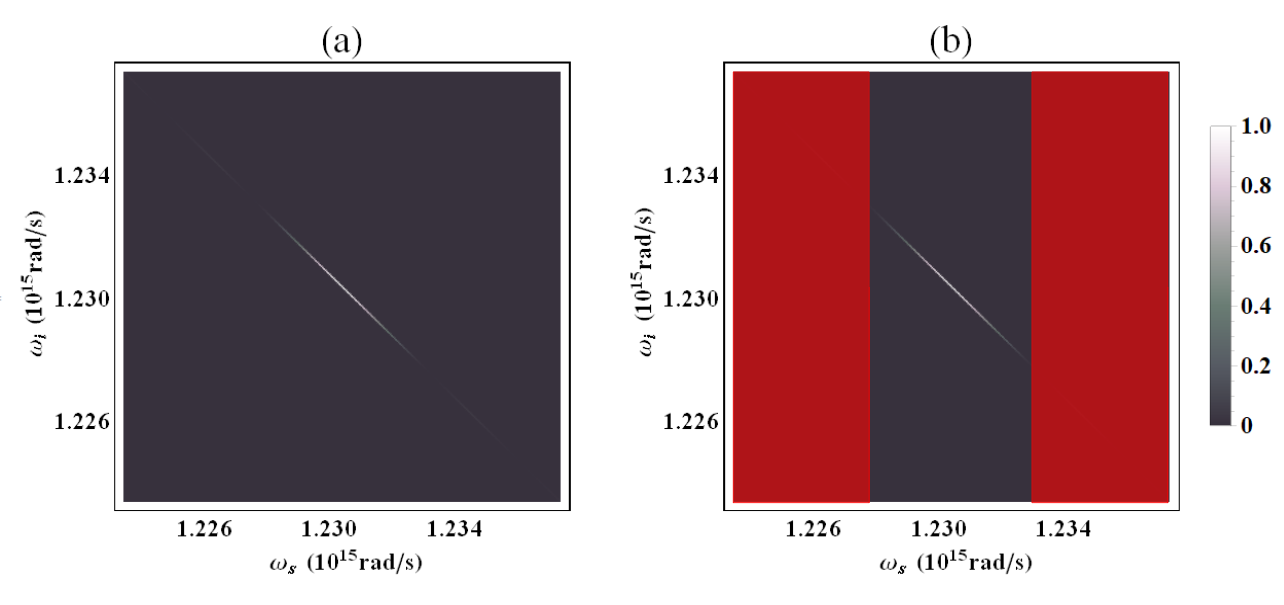}
	\caption{The normalized JSI at $\phi=0$: (a) without filter, (b) with the filter (bandwidth 5.71$\cdot 10^{12}$ rad/s).}
	\label{filter}
\end{figure*}


Fig.\ref{psvsgain} shows the minimum values of the normalized phase sensitivity for each gain.  It is clearly seen that the phase sensitivity gets worse with enhancing gain or pump power. Explicit calculations of the mean photon number and its variance can be found in the Appendix, where it is shown that the main trend of the normalized phase sensitivity with increasing gain is roughly defined as:
\begin{equation}
\frac{\Delta\phi}{\Delta\phi_{SNL}}\approx\frac{ \sinh[\gamma/2]}{\gamma}.
\label{ps_gain}
\end{equation}
This trend can be observed in Fig.\ref{psvsgain} and reflects a progressive worsening of $\Delta\phi$ as the gain increases. However, the phase sensitivity can be improved by using a filtering of the output radiation or a seed, both strategies will be investigated below.
\section{Filtering}
In this section, we demonstrate how to improve the phase sensitivity of the SU (1,1) interferometer by filtering a specific frequency range in the output radiation.

Since filters select specific spectral bandwidths, working in the Schmidt basis is no longer meaningful as the Schmidt operators include the integration over all frequencies. Hence, it is more convenient to calculate the relevant quantities by using the output annihilation and creation operators expressed in the plane wave basis, introduced in \cite{PhysRevA.91.043816} and directly presented in Appendix, see Eq.(\ref{ab}). Using the output plane wave operators Eq.(\ref{ab}), the number of signal photons in a certain spectral range $2 \delta$ around the central frequency $\omega_p/2$ is given by:
\begin{widetext}
\begin{equation}\begin{split}
\langle N_s(\phi)\rangle=\int_{\omega_p/2-\delta}^{\omega_p/2+\delta}d\omega_s\langle a_s^{\dagger out} (\omega_s)a_s^{out}(\omega_s)\rangle
=\sum_k\sinh^2\gamma_k\int_{\omega_p/2-\delta}^{\omega_p/2+\delta}d\omega_s \lvert u_k(\omega_s,\phi)\rvert^2 ,
\label{spectr}
\end{split}\end{equation}
\end{widetext}
where $\delta$ determines the filter bandwidth and $\gamma_k=G(\phi)\sqrt{\lambda_k(\phi)}$, see Appendix. 
Similarly, we can calculate the variance of the number of photons:
\begin{widetext}
\begin{equation}\begin{split}
\langle\Delta^2 N_{s}(\phi)\rangle=\langle N_{s}(\phi)\rangle
+\sum_{kk'}\left\lvert\int_{\omega_p/2-\delta}^{\omega_p/2+\delta}d\omega_su_k^*(\omega_s,\phi)u_{k'}(\omega_s,\phi) \sinh\gamma_k\sinh\gamma_{k'}\right\rvert^2,
\label{disp}
\end{split}\end{equation}
\end{widetext}

where the first term corresponds to Eq.(\ref{spectr}).  Using both the variance in Eq.(\ref{disp}) and the derivative of Eq.(\ref{spectr}), the phase sensitivity in a filtering case can be calculated according to Eq.(\ref{psdef}).

It is important to specify that considered filters should also be taken into account when calculating the SNL via Eq.(\ref{snl}). In this way, the number of photons in the SNL corresponds to the number of signal photons  generated by a single PDC section and subject to the band-pass filter, these photons then enter the phase modulator:
\begin{equation}\begin{split}
\langle N_{in}\rangle=\sum_k\sinh^2\left[ G_1\sqrt{\eta_k}\right]\int_{\omega_p/2-\delta}^{\omega_p/2+\delta}d\omega_s \lvert \bar u_k(\omega_s)\rvert^2 ,
\label{SNL}
\end{split}\end{equation}
where $\bar u_k(\omega_s)$ is the orthonormal set of signal Schmidt modes corresponding to a single periodic poled waveguide, both $G_1$ and  $\eta_k$ are defined in the Appendix.

The filter we used selects a range with bandwidth $2\delta=5.71 \cdot 10^{12} rad/s$ around the central frequency, see Fig.(\ref{filter}). This filter cuts the main part of the JSA with the central peak, excluding the second order peaks of the sinc function in Eq.(\ref{JSAcomp2}), and allows us to investigate the influence of the side-lobes of the JSA on the phase sensitivity.
\begin{figure*}
	\centering
	\includegraphics[width=0.79\linewidth]{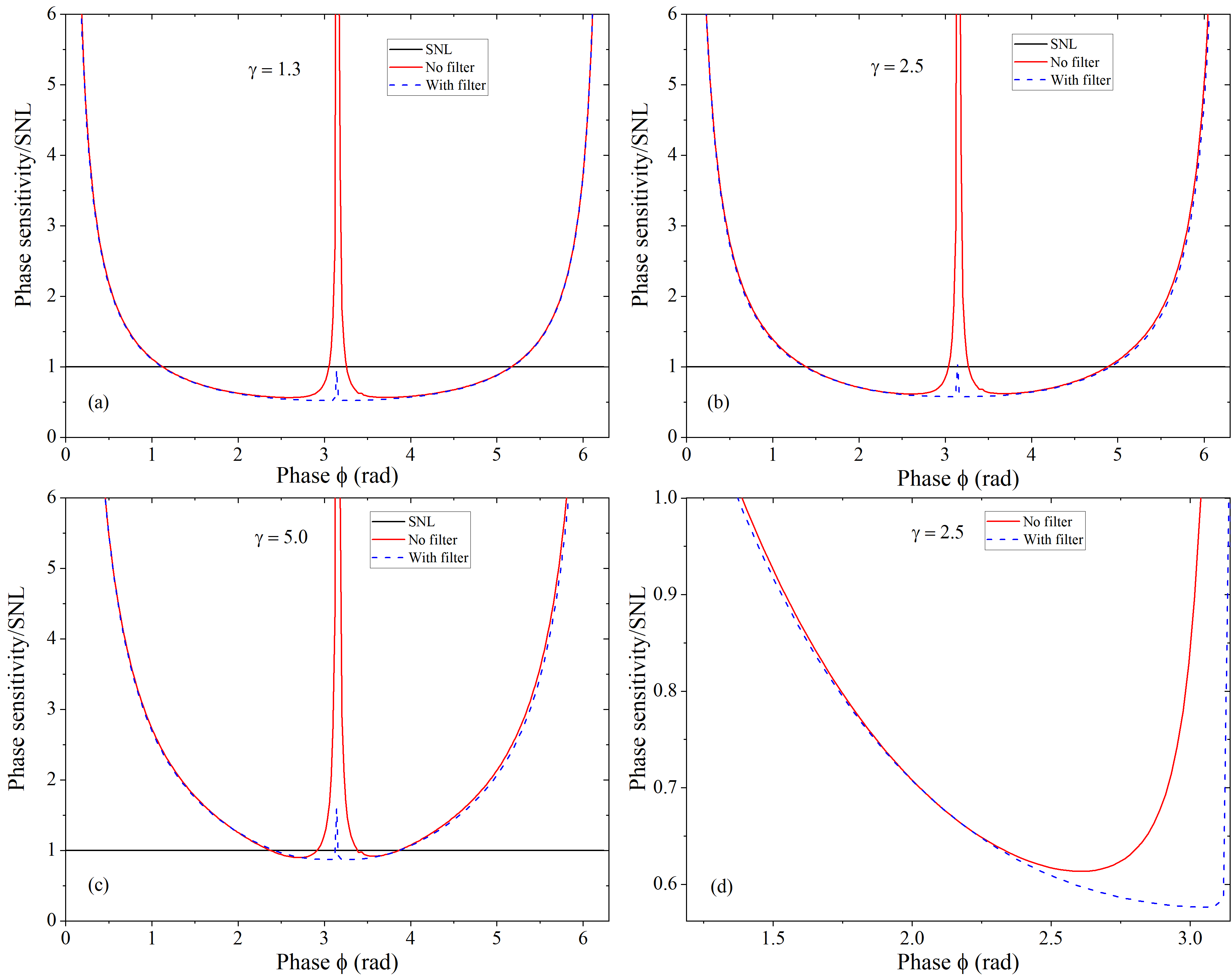}
	\caption{The normalized phase sensitivity with and without a filter for different gains: (a) $\gamma=1.3$,  (b) $\gamma=2.5$ and (c) $\gamma=5.0$. (d) The zoom of the supersensitivity region for $\gamma=2.5$ . The solid red line shows the case without a filter, the dashed blue line presents the filter case.}
	\label{psfilt}
\end{figure*}
The normalized phase sensitivity in the filtered case at different gains is presented in Fig.\ref{psfilt}.
It can be seen that in the region far from $\phi=\pi$, the filter provides similar results to the no-filter case. However, around $ \phi = \pi $, filtering gives a significant improvement in the phase sensitivity both in depth and in the expansion  of the supersensitivity region, highlighted in Fig.\ref{psfilt}d. 
This improvement indicates the negative influence of the side lobes. Indeed, removing the side lobes does not dramatically change the final number of photons with respect to the case without a filter, nevertheless, it reduces the influence of the residual photons stemming from the non-perfect compensation in JSA for large frequency detuning $\lvert\Omega\rvert>0$ at $\phi\simeq\pi$.
\begin{figure}[H]
	\centering
	\includegraphics[width=1\linewidth]{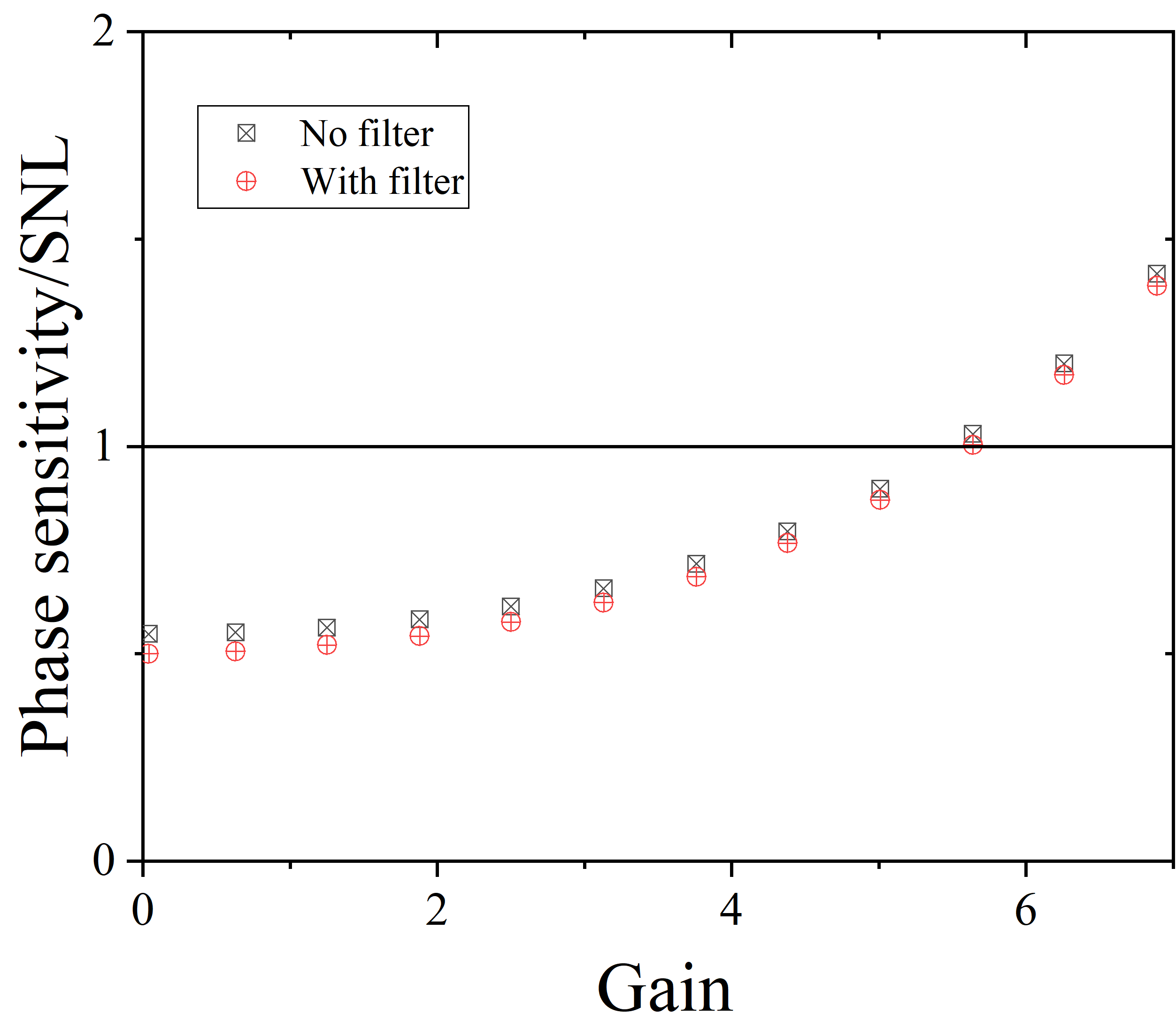}
	\caption{The minimum values of the normalized phase sensitivity versus gain $\gamma$ with and without the use of the filter. The SNL is represented by a solid black line.}
	\label{pgfilter}
\end{figure}
The best phase sensitivity achieved with and without the filter at varying gains is shown in Fig.\ref{pgfilter}. 
This picture demonstrates that the action of the filter guarantees better phase sensitivity with respect to the no-filter case, but at the same time confirms the tendency of worsening the phase sensitivity with gain as in the case without a filter.

\section{Seeded SU(1,1)}\label{seeding}
The description of the SU(1,1) interferometer in terms of broadband spectral modes reveals the possibility to prepare the seed either at the selected frequency or at the selected spectral mode of the interferometer (or their superposition). In this section, we analyze the seeding of the interferometer by a single photon in the first Schmidt mode. Other cases are analysed in the Appenix \ref{app2}. The seed is performed only for the signal mode of the interferometer (the vacuum input for the  idler mode), which can be done due to the distinguishability of photons because of their different polarizations. Since the Schmidt modes of the SU(1,1) interferometer depend on the internal phase, we suppose to utilise the radiation generated by the second interferometer to prepare a seed in the selected phase-dependent Schmidt mode.

\subsection{Direct detection}
When seeding, the detection strategy is important for the phase sensitivity. In this subsection we consider direct detection, as was done in the previous section without seeding. Therefore, the protocol as well as the set of operators we need in order to estimate the phase sensitivity in Eq.(\ref{psdef}) are the same as before, but the initial states are different.

\subsubsection{Single photon in the first Schmidt mode} 
\begin{figure}[H]
	\centering
	\includegraphics[width=1\linewidth]{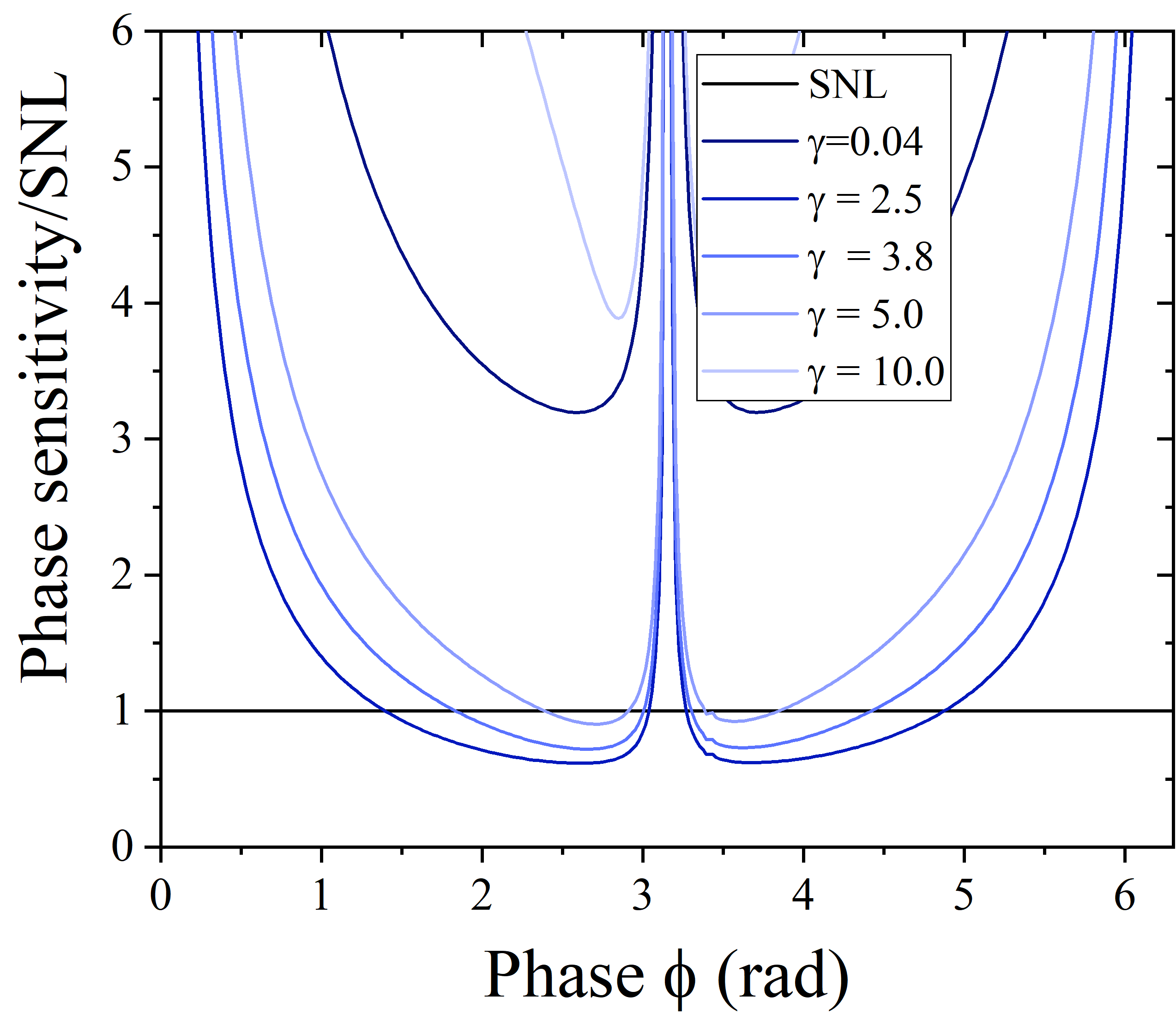}
	\caption{The normalized phase sensitivity on the phase for different gains. The first Schmidt mode of the signal photon is seeded with a single photon. The SNL is represented by a solid black line. }
	\label{s1p}
\end{figure}

\begin{figure}[H]
	\centering
	\includegraphics[width=1\linewidth]{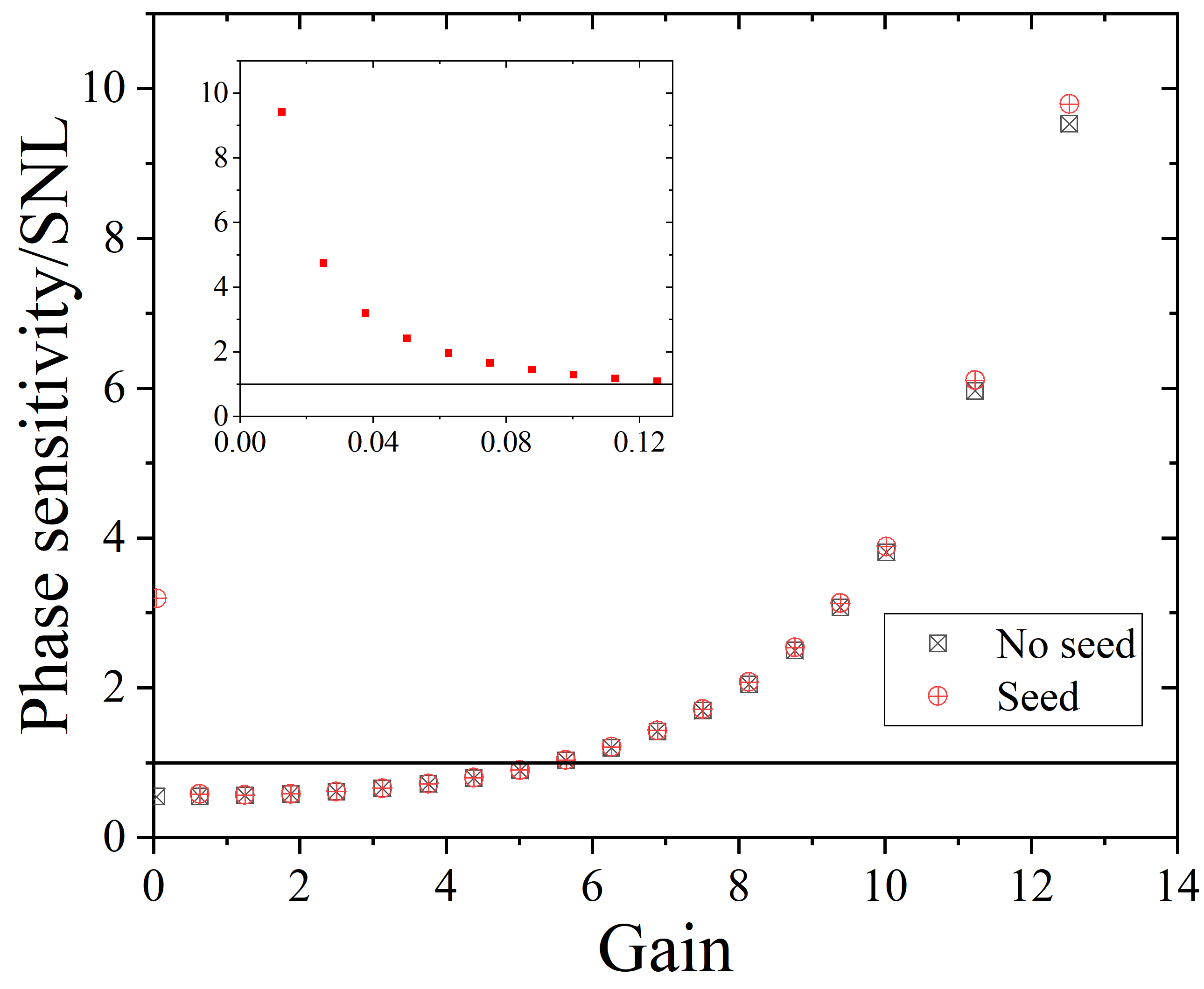}
	\caption{The phase sensitivity trend in Fig.\ref{s1p} at the point of minimum with respect to the gain (red circles) compared to the no-seeding case (black squares). The SNL is plotted in black. The insert shows the zoom of the phase sensitivity in the seeded case for small gain values.}
	\label{seed1}
\end{figure}

Having a single photon in the first Schmidt mode, the input state of the interferometer is given by:
\begin{equation}
\vert\psi\rangle_{in}= \vert 1\rangle_{A_1}\vert 0\rangle_{A_{n\neq 1}}\vert 0\rangle_{B_n}.
\end{equation}
This state shows that only the first Schmidt mode of the signal photon is seeded, while the vacuum stands for other input signal Schmidt modes and all idler  modes.
Direct calculations show that the output number of signal photons in this configuration is:
\begin{equation}
\langle N_s \rangle=\sum_k \sinh^2\gamma_k+1+\sinh^2 \gamma_1.
\label{Nsvac}
\end{equation}
The first term in Eq.(\ref{Nsvac}) is the vacuum term obtained without a seed and already known in Eq.(\ref{np}). The second and the third terms appear because of the seed: the unity term describes the single photon, whereas the last term is the surplus of photons in the signal mode generated by the seed.
The variance of the number of photons is given by:
\begin{equation}
\langle\Delta^2 N_s\rangle=\sum_k\sinh^2 \gamma_k\cosh^2 \gamma_k+\sinh^2 \gamma_1\cosh^2 \gamma_1.
\end{equation}
The normalized phase sensitivity as a function of phase for different gains is plotted in Fig.\ref{s1p}. Fig.\ref{seed1} presents the best phase sensitivity for each gain. Overall, this figure shows a similar behaviour with respect to the case with a vacuum seed, except for small gains when the number of PDC photons is less than one: for this region, the phase sensitivity becomes drastically worse. 

This is due to the fact that the shot noise limit in the case of the single-photon-seed
\begin{equation}\begin{split}
\Delta\phi_{SNL}=\frac{1}{\sqrt{1+\sinh^2[C_1\sqrt{\eta_1}]+\sum_k \sinh^2[C_1\sqrt{\eta_k}]}}
\end{split}\end{equation}
tends to unity for small gains, or in other words, only the seeded photon contributes to SNL; while, the non-normalized phase sensitivity is still proportional to $1/\Gamma$. 
For gain values greater than unity, the second and third terms dominate in the SNL, and the phase sensitivity grows exponentially, similarly to the case without a seed, see Fig.(\ref{seed1}). In other words, the single-photon seed leads to an imbalance in the signal-idler photon correlations for low gains, which destroys the phase sensitivity in this region. The analysis of other single-mode seeding cases and detection stratagies shows that such inefficiency of a single-mode seed is an intrinsic property of a multimode interferometer that fundamentally distinguishes it from a single-mode system.

\section{Conclusion}
In this work, we presented a detailed theoretical description of a fully integrated multimode SU(1,1) interferometer. We demonstrated that material dispersion complicates the interferometer setup and special integrated designs are required. The presented interferometer includes a polarization converter located between the two photon sources, which allows us to compensate for the time delay arising between signal and idler photons.  The use of the CW laser allows us to avoid additional time compensation for the pump photons and, at the same time, to  reduce the influence of uncompensated second-order effects, that results in the supersensitivity regions below the shot noise limit. The spectral properties of such interferometer were investigated and appropriate conditions for the best phase sensitivity below the shot noise limit were found and analysed. Interestingly, the presented configuration leads to a situation where the multimode system realized with  using a CW laser outperforms the single-Schmidt-mode system realized with a pulsed laser.

In addition, we investigated the features of the phase sensitivity with an increase in the gain of the process and show the degradation of the phase sensitivity with increasing gain due to the multimode structure of generated light.

Finally, we discuss the impact of filtering and seeding on the phase sensitivity. We show that using filtering one can remove the uncompensated radiation and improve the phase sensitivity both in depth and in the expansion of the supersensitivity regions. However, the investigated seeds in the signal mode do not improve the phase sensitivity and even completely destroy the supersensitivity regions in the case of a bright single-mode seed. The reason is the destruction of signal-idler photon correlations, which are strongly required for the observation of supersensitivity in the highly multimode regime.

This work stresses the fundamental difference between the single-plane-wave-mode and strongly multimode interferometers in the behaviour of the phase sensitivity.
Taking into account the spectral features of photon sources is an important point for more pragmatic descriptions of this type of interferometer and future experimental implementations and investigations.
Moreover, this work demonstrates that the introduction of spectral modes in the phase sensitivity analysis could open new frontiers in quantum metrology: it was shown how an accurate spectral engineering in terms of mode scaling and frequency filtering determines deep changes in final results; furthermore, a large spectrum of seeding strategies can be explored.
Finally, the impact of internal losses on the performance of the multimode integrated nonlinear interferometer is an important open question that will be a scope for future research.

\section{ACKNOWLEDGMENTS}
We acknowledge the financial support of the Deutsche Forschungsgemeinschaft (DFG)
via TRR~142/2, project C02.  P.~R.~Sh. thanks the state of North Rhine-Westphalia for support by the
{\it Landesprogramm f{\"u}r geschlechtergerechte Hochschulen}. 
We also thank the $\mathrm{PC}^2$ (Paderborn Center for Parallel Computing) for providing computing time.

\printbibliography
\appendix
\section{Schmidt modes calculation}\label{app1}
The PDC process is a well-known source of broadband squeezed light. In order to make the squeezing and broadband behaviour mathematically explicit, we firstly need to introduce the Schmidt operators, namely a special set of annihilation and creation operators equipped with a spectral distribution. For doing this, we should use the Schmidt decomposition of the JSA presented in Eq.(\ref{jsa}), 
\begin{equation}
F(\omega_s,\omega_i,\phi)=\sum_k \sqrt{\lambda_k(\phi)} u_k(\omega_s,\phi)v_k(\omega_i,\phi),
\end{equation}
where we made explicit the parametric dependence of the JSA on the phase  $\phi$ experienced by the idler photon. In this framework, $\lambda_k(\phi)$ are the Schmidt eigenvalues and $u_k(\omega_s,\phi)$ and $v_k(\omega_i,\phi)$ are an orthonormal set of spectral Schmidt modes for signal and idler photons, respectively. At this point it is convenient to introduce the input Schmidt operators as following:
\begin{equation}\begin{split}
(A_k^\dagger)^{in}=\int d\omega_s u_k(\omega_s,\phi)a^\dagger(\omega_s),\\
(B_k^\dagger)^{in}=\int d\omega_i v_k(\omega_i,\phi)a^\dagger(\omega_i),
\label{oper}
\end{split}\end{equation}
which describe the signal and idler photons characterized by the spectra $\lvert u_k(\omega_s,\phi)\rvert^2$ and $\lvert v_k(\omega_i,\phi) \rvert^2$, respectively. These operators fulfill the typical bosonic commutation rules:
\begin{equation}
[A_k,A^\dagger_j]=\delta_{kj}, \, [A_k,B^\dagger_j]=0.
\label{comm}
\end{equation}
Following the strategy detailed in \cite{PhysRevA.91.043816}, we can define the Hamiltonian in terms of Schmidt operators and solve the Heisenberg equations in order to find both the output Schmidt operators:
\begin{equation}\begin{split}
A_k^{out}=A_k^{in}\cosh\left[G(\phi) \sqrt{\lambda_k(\phi)}\right]+\\(B_k^\dagger)^{in}\sinh\left[G(\phi) \sqrt{\lambda_k(\phi)}\right],\\
B_k^{out}=B_k^{in}\cosh\left[G(\phi) \sqrt{\lambda_k(\phi)}\right]+\\(A_k^\dagger)^{in}\sinh\left[G(\phi) \sqrt{\lambda_k(\phi)}\right],
\label{AB}
\end{split}\end{equation}
and the output plane wave operators:
\begin{equation}
\begin{split}
a^{out}(\omega_s)= a^{in}(\omega_s)+ \sum_k u_k(\omega_s,\phi)\\
+\bigg[A_k^{in}\left(\cosh\left[G(\phi) \sqrt{\lambda_k(\phi)}\right]-1\right)\\+(B_k^\dagger)^{in}\sinh\left[G(\phi) \sqrt{\lambda_k(\phi)}\right]\bigg],\\ \\
b^{out}(\omega_i)=b^{in}(\omega_i)+ \sum_k v_k(\omega_i,\phi)\\
+ \bigg[B_k^{in}\left(\cosh\left[G(\phi) \sqrt{\lambda_k(\phi)}\right]-1\right)\\
+(A_k^\dagger)^{in}\sinh\left[G(\phi) \sqrt{\lambda_k(\phi)}\right]\bigg],
\label{ab}
\end{split}
\end{equation}
where $G(\phi)=\int C(\phi)\Gamma dt$, $C(\phi)$ is the normalization constant depending on the phase, see Eq. \ref{jsa}, the coupling constant $\Gamma$ is proportional to the pump power. 

The output operators can be used to calculate different quantities, for instance,  the output integral mean number of signal photons is given by:
\begin{equation}\begin{split}
\langle N_s(\phi)\rangle=\sum_k \sinh^2\left[ G(\phi)\sqrt{\lambda_k(\phi)}\right].
\label{np}
\end{split}\end{equation}
Since we only consider the signal photon observables, the index "s" will be omitted below along the text.

Similarly, the variance of the integral number of signal photons is given by:
\begin{equation}
\langle\Delta^2 N\rangle=\frac{1}{4}\sum_k \sinh^2\left[2G(\phi)\sqrt{\lambda_k(\phi)}\right].
\label{delta}
\end{equation}
The derivative of the mean number of photons $\langle N\rangle$ respect to the phase gives 
\begin{equation}\begin{split}
\frac{d\langle N\rangle}{d\phi}=\sum_k \sinh\left[2 G(\phi)\sqrt{\lambda_k(\phi)}\right] \\
\times	\left(\frac{d G(\phi)}{d\phi}\sqrt{\lambda_k(\phi)}+\frac{G(\phi)}{2\sqrt{\lambda_k(\phi)}}\frac{d\lambda_k(\phi)}{d\phi}\right).
\label{der}
\end{split}\end{equation}

In all cases analysed in this work, the JSA never vanishes totally due to the multimode structure of the light. This means that both $G(\phi)$ and the eigenvalues $\lambda_k(\phi)$ cannot reach zero, as a result,  both the mean photon number and its variance are not fully suppressed. Therefore, by the definition of phase sensitivity in Eq.\ref{psdef}, the non vanishing variance causes the presence of a peak in the phase sensitivity plot. 

It is fundamental to assert that the plotted phase sensitivity is normalized with respect to the SNL, which depends on the number of signal photons inside the interferometer:
\begin{equation}
\langle N_{in} \rangle=\sum_k \sinh^2\left[G_1\sqrt{\eta_k}\right],
\label{npsnl}
\end{equation}
where $\eta_k$ are the Schmidt eigenvalues corresponding to the single periodically poled waveguide.  $G_1=\int C_1 (\Gamma/2) dt$,  where it is taken into account that the coupling constant of a single PDC section is twice smaller compared to the double-PDC-section configuration; $C_1$ is the normalization constant corresponding to the JSA of a single PDC section. 

In order to give an analytical justification of the trend of the normalized phase sensitivity in Fig. \ref{psvsgain}, it is more practical to split our argumentation in two parts, distinguishing a low gain regime, in which $G(0)\sqrt{\lambda_1(0)}\ll 1$, from a high gain regime, in which $G(0)\sqrt{\lambda_1(0)}\gg 1$.

In the former case, an estimation of the photon number in Eq.(\ref{np}) and in Eq.(\ref{npsnl})  gives
\begin{equation}
\langle N(\phi)\rangle\simeq G^2(\phi) \sum_k \lambda_k(\phi) = G^2(\phi),
\label{np2}
\end{equation}
and 
\begin{equation}
\langle N_{in}\rangle\simeq G_1^2\sum_k \eta_k=G_1^2,
\end{equation}
respectively, whereas using the Taylor expansion of the variance Eq.(\ref{delta})  we obtain
\begin{equation}
\langle\Delta^2 N\rangle\simeq G^2(\phi)\sum_k \lambda_k(\phi) =G^2(\phi).
\end{equation}
Since the SU(1,1) interferometer, which we are dealing with, has the JSA at $\phi=0$ almost identical to the JSA of a single waveguide, we can assume that $C_1\approx C(0)$ and $\lambda_k(0)  \approx \eta_k$.
Therefore, by calculating the derivative of the photon number in Eq.(\ref{np2}) with respect to the phase, it turns out that the normalized phase sensitivity is described by:
\begin{equation}
\frac{\langle \Delta\phi\rangle}{\langle \Delta\phi_{SNL}\rangle}\approx\left\lvert\frac{G(0)}{4}\left(\frac{\partial G(\phi)}{\partial \phi}\right)^{-1}\right\rvert\approx \frac{1}{2\sin(\phi/2)},
\end{equation}
where the last step stems from the assumption that in the low intensity regime, according to the form of JSA Eq.(\ref{JSAcomp2}),  the gain is a modulated function of the form $G(\phi)=G(0)\lvert\cos(\phi/2)\rvert$ (this expression holds at $0<\phi<\pi$). It is interesting to notice that the normalized phase sensitivity in the low gain regime does not depend on the gain anymore, which can be seen in Fig. \ref{psvsgain}.

\ifx 
In the high gain regime, the intensity of the pump laser causes a redistribution of $\lambda_k(\phi)$, and the first modes tends to contribute more to the average number of photons in Eq. (\ref{np}) \cite{PhysRevA.97.053827}. It is possible to estimate the effective number of modes contributing to Schmidt number $K = 1/(\sum_k \Lambda_k^2)$, where 
\begin{equation}
\Lambda_k=\frac{\sinh^2\left[ G(\phi)\sqrt{\lambda_k(\phi)}\right]}{\sum_k \sinh^2\left[ G(\phi)\sqrt{\lambda_k(\phi)}\right]}
\end{equation}
is a new set of coefficients, 
Assuming therefore 
$\langle N(\phi)\rangle\approx \overbar K  \sinh^2\left[ G(\phi)\sqrt{\lambda_{\overbar K}(\phi)}\right]$
\fi
In the high gain regime, due to the redistribution of the eigenvalues  \cite{PhysRevA.91.043816}, the contribution of the first mode to the signal and variance will be dominated, and we can temporarily underestimate the number of photons of both equations (\ref{np}) and (\ref{npsnl}) by taking into account the contribution stemming from the first mode, obtaining:
\begin{equation}
\langle N_s(\phi)\rangle\approx\sinh^2\left[ G(\phi)\sqrt{\lambda_1(\phi)}\right],
\end{equation}
and
\begin{equation}
\langle N_{in} \rangle\approx \sinh^2\left[G_1\sqrt{\eta_1}\right],
\end{equation}
respectively. We underestimate the variance consequently:
\begin{equation}
\langle\Delta^2 N\rangle\approx\frac{1}{4} \sinh^2\left[2G(\phi)\sqrt{\lambda_1(\phi)}\right].
\end{equation}
By assuming again $C_1\approx C(0)$ and $\lambda_k(0)  \approx \eta_k$, the gain of the process can also be expressed as $\gamma = G(0)\sqrt{\lambda_1(0)}\simeq 2G_1\sqrt{\eta_1}$. Moreover, the shape of the JSA  Eq.(\ref{JSAcomp2}) modulated by $\cos(\phi/2)$ allows us to write $G(\phi)\sqrt{\lambda_1(\phi)}\approx \gamma \lvert\cos(\phi/2)\rvert$. Finally, by collecting all assumptions written  above, we can estimate the normalized phase sensitivity:
\begin{equation}
\frac{\Delta\phi}{\Delta\phi_{SNL}}\approx \frac{\sinh[\gamma/2]}{\gamma\sin(\phi/2)}.
\label{ps}
\end{equation}

If the angle $\phi$ is close to $\pi$ (the point of almost perfect destructive interference), the eigenvalues distribution is very close to the flat distribution. In this case, the underestimation with only the first Schmidt mode is not correct, but one can use the overestimation, suggesting that all Schmidt modes have the same weights. It means that the integral number of photons is equivalent to
\begin{equation}
 \langle N(\phi)\rangle\approx K  \sinh^2\left[ G(\phi)\sqrt{\lambda_{ 1}(\phi)}\right],
 \label{nps}
 \end{equation}
where we assume that the all $K$ Schmidt modes have the same contribution,  $K = 1/(\sum_k \Lambda_k^2)$ is the effective number of modes (the Schmidt number) and
\begin{equation}
\Lambda_k=\frac{\sinh^2\left[ G(\phi)\sqrt{\lambda_k(\phi)}\right]}{\sum_k \sinh^2\left[ G(\phi)\sqrt{\lambda_k(\phi)}\right]}
\end{equation}
is a set of Schmidt coefficients taking into account the redistribution of the weights in the high gain regime \cite{PhysRevA.97.053827}. 
The variance would change consequently:
\begin{equation}
\langle\Delta^2 N\rangle=\frac{K}{4} \sinh^2\left[2G(\phi)\sqrt{\lambda_1(\phi)}\right].
\end{equation}
By modifying the mean photon number in the SNL similarly to Eq.(\ref{nps}), along with all assumptions made above, the overestimated normalized phase sensitivity would be identical to Eq.(\ref{ps}), which in turn, with $\phi=\pi$ equals to Eq.(\ref{ps_gain}).

\section{Further seeding configurations}\label{app2}
In this appendix, we analyse different types of seeding as well as different detection strategies to find the optimal conditions for the best phase sensitivity. In all cases presented below, the seed consists of a coherent state, moreover, seeding is performed only for the signal mode of the interferometer (the vacuum input for the  idler mode), which can be done due to the distinguishability of photons because of their polarizations.
The seed of coherent states in specific modes can be carried out experimentally by injecting a laser whose spectral features are shaped to one of the specific modes of the interferometer. Furthermore, since the Schmidt modes of the interferometer depend on the internal phase, we can suppose to tune the spectral profile of the laser accordingly.
\subsection{Direct detection: coherent state in the first Schmidt mode} 
In this configuration, we utilise an intense coherent state $\vert \alpha\rangle$ in the first Schmidt mode of the signal photon, the input state in this case is given by:
\begin{equation}
\vert\psi\rangle_{in}= \vert \alpha\rangle_{A_1}\vert 0\rangle_{A_{n\neq 1}}\vert 0\rangle_{B_n}.
\label{seedalpha}
\end{equation}
To simplify calculations, we consider a coherent state with a real $\alpha$ and high number of photons $\lvert \alpha \rvert^2=10^6$.  Such seeding changes the SNL accordingly:
\begin{equation}\begin{split}
\Delta\phi_{SNL}=\\
\frac{1}{\sqrt{\alpha^2\cosh^2[C_1\sqrt{\eta_1}]+\sum_k \sinh^2[C_1\sqrt{\eta_k}]}}.
\label{SNLalpha}
\end{split}\end{equation}
Analogous to the previous case, direct calculations give the following expressions for the mean photon number:
\begin{equation}
\langle N_s \rangle=\sum_k \sinh^2\gamma_k+\vert\alpha\vert^2+\vert\alpha\vert^2\sinh^2 \gamma_1,
\end{equation}
and for the photon number variance:
\begin{equation}\begin{split}
\langle\Delta^2 N_s\rangle=\sum_k\sinh^2 \gamma_k\cosh^2 \gamma_k\\
+\vert\alpha\vert^2\cosh^2 \gamma_1\cosh 2\gamma_1.
\end{split}\end{equation}

\begin{figure}[H]
	\centering
	\includegraphics[width=1\linewidth]{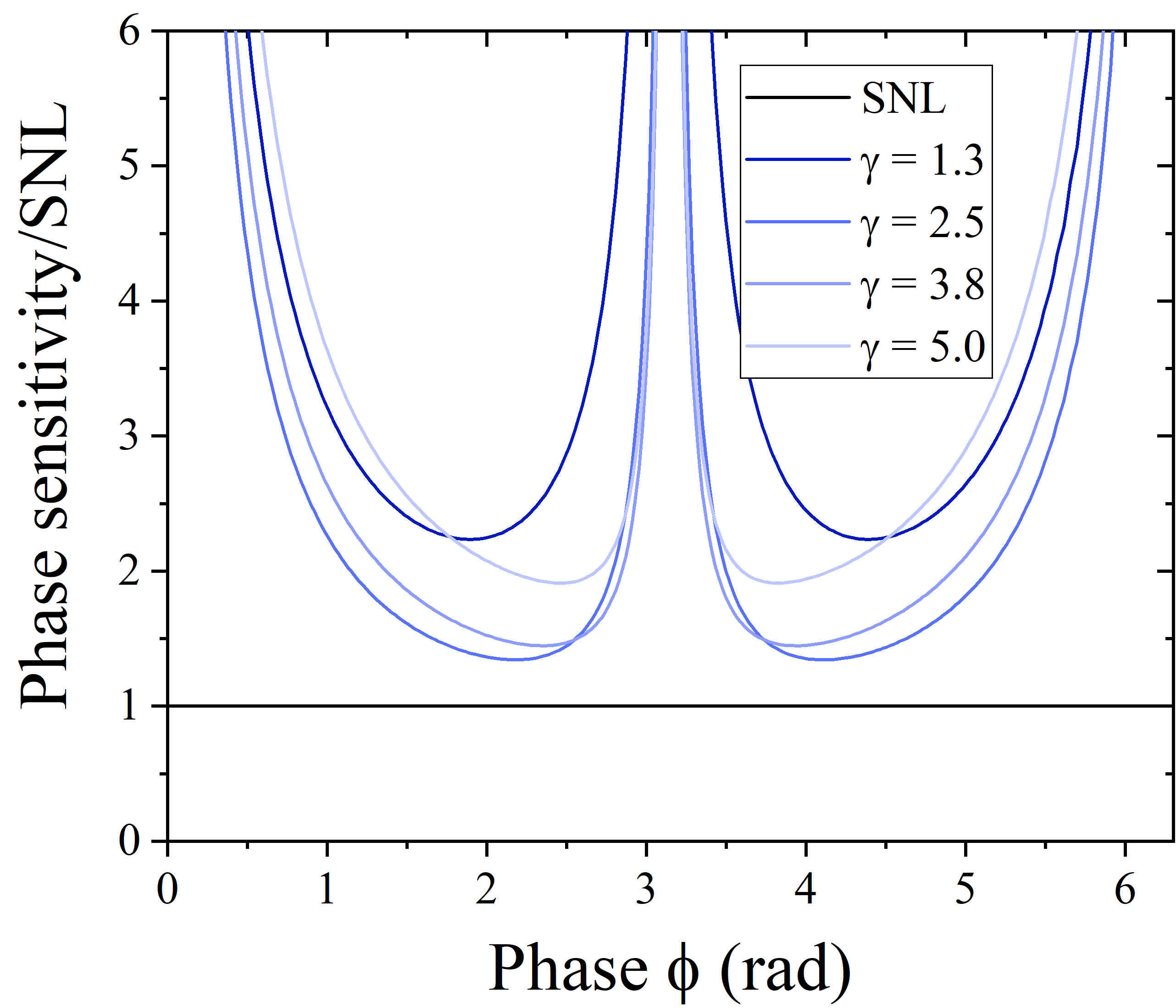}
	\caption{The normalized phase sensitivity on the phase for different gains.  The first Schmidt mode of the signal photon is seeded with an intense coherent state having $\lvert \alpha\rvert^2=10^6$ photons. The SNL is represented by a solid black line. }
	\label{salphap}
\end{figure}

However, as it is shown in Fig.\ref{salphap}, for any values of the gain this configuration does not present supersensitivity regions. Analytically, this can be demonstrated by checking the normalized phase sensitivity in the asymptotic limit when $\alpha^2\gg\sum_k \sinh^2\gamma_k$, namely
\begin{equation}
\frac{\Delta^2\phi}{\Delta^2\phi_{SNL}}\approx(1+\coth^2\gamma_1)\cosh^2\left[G_1\sqrt{\eta_1}\right]\left\lvert\frac{\partial\gamma_1}{\partial\phi}\right\rvert^{-2}. 
\end{equation}
Due to the similarity between the JSA in Eq.(\ref{JSAcomp2}) at $\phi=0$ and the JSA of a single waveguide, we can assume $\lambda_1(0)\simeq\eta_1$ and $G(0)\simeq 2 G_1$; hence the gain $\gamma$ can be expressed as $\gamma\simeq 2G_1\sqrt{\eta_1}$. Furthermore, we can also assume $\gamma_1\approx\gamma\lvert\cos(\phi/2)\rvert$, therefore in the range $\{0-\pi\}$ we have
\begin{equation}
\frac{\Delta^2\phi}{\Delta^2\phi_{SNL}}\approx\frac{4\left(1+\coth^2[\gamma\cos(\phi/2)]\right)\cosh^2[\gamma/2]}{\gamma^2\sin^2(\phi/2)}. 
\end{equation}
By plotting this curve, or alternatively by calculating the minimum point with respect to $\phi$, it is clear that this function cannot even reach the shot noise limit.

\subsection{Homodyne detection}
In this subsection we discuss seeded interferometers with the homodyne detection strategy. In the case of homodyne detection, we must properly modify  the definition of the phase sensitivity using  the homodyne operator $\hat  H_d$ (not the photon number operator as before) \cite{PhysRevA.95.063843} to calculate observables. According to this modification, the phase sensitivity is defined as:
\begin{equation}
\Delta\phi=\bigg\lvert\frac{\langle\Delta \hat H_d\rangle}{\partial \langle \hat H_d \rangle/\partial\phi}\bigg\rvert.
\label{pshom}
\end{equation}
The homodyne operator depends on the local oscillator properties, which is why it is necessary to ensure that the output light of the SU(1,1) interferometer and the local oscillator are spectrally matched.
To ensure this, both the output radiation and the local oscillator are spectrally filtered. The spectral characteristic of the filters will be modelled depending on the different type of seeding used. In particular, in case 1 below, we will consider a spectral filter selecting only the first Schmidt mode, while in case 2 we will consider a passband filter around the central frequency of generated squeezed light. The SNL calculation takes these modifications into account as well: the mean number of photons in the SNL corresponds to the number of signal photons generated by a single periodically poled waveguide, averaged with respect to the input seeding state, and then filtered to match the spectral properties of the output squeezed light.
\subsubsection{Coherent state in the first Schmidt mode}
The first case we consider is an intense coherent seed in the first Schmidt mode with the same initial state as in Eq.(\ref{seedalpha}). At the end of the interferometer we filter the output radiation of the signal photon, selecting the first Schmidt mode. A local oscillator is fixed in the same frequency range, corresponding to the first Schmidt mode. 
In this case, the homodyne operator is defined as:
\begin{equation}
\hat H_d=\lvert\beta_{lo}\rvert\left(e^{i\theta_a}\hat  A_1^{out}+e^{-i\theta_a}\left[\hat  A_1^{out}\right]^\dagger\right),
\end{equation}
where $\lvert\beta_{lo}\rvert$ and $\theta_a$ are the amplitude and the phase of the local oscillator (described by the coherent state $\vert\lvert\beta_{lo}\rvert e^{i\theta_a}\rangle_{A_1}$).
Averaging the homodyne operator with respect to the input state we obtain:
\begin{equation}
\langle 
\hat  H_d\rangle=2 \alpha \lvert\beta_{lo}\rvert\cos\theta_a\cosh\gamma_1.
\label{hom}
\end{equation}
The variance of the homodyne operator is given by:
\begin{equation}
\langle 
\Delta^2 \hat H_d\rangle=\lvert\beta_{lo}\rvert^2\cosh{2\gamma_1},
\label{deltahom}
\end{equation}
whereas the expression for the shot noise limit is:
\begin{equation}
\Delta\phi_{SNL}=\frac{1}{\sqrt{\alpha^2\cosh^2[C_1\sqrt{\eta_1}]+ \sinh^2[C_1\sqrt{\eta_1}]}}.
\end{equation}
The optimum phase sensitivity is reached when $\theta_a=0$.
\ifx
Here we notice that the trend of the phase sensitivity in relation with the phase does not differ significantly respect to what was observed with the same seeding but direct detection. Also in this case, a broadening supersensitivity range is observed for higher gain, and this behaviour holds as long as the amount of seeding photons is higher than the photons generated by the PDC crystal, as it is shown in Fig. \ref{alphahom}, where a comparison with the direct detection is represented. 
Mathematically, this similarity is immediately understandable by looking at Eq.(\ref{psalpha}): this is the exact expression we would get by dividing the square root of Eq.(\ref{deltahom}) by 
\begin{equation}
\left\lvert\frac{\partial H_q}{\partial\phi}\right\rvert=\lvert\beta_{lo}\rvert\sinh{\gamma_1}\left\lvert\frac{\partial \gamma_1}{\partial\phi}\right\rvert,
\end{equation}
namely the derivative of Eq.(\ref{hom}).

However, a discrepancy for higher gain emerges: the filter used in the homodyne detection selects only photons in the first mode, whereas light from other modes is blocked. This effect must be taken into account in order to estimate the SNL, since we should suppose to filter the radiation leaving the classical SU(2) interferometer accordingly. Hence, the expression of the SNL corresponds here to:
\begin{equation}
\Delta\phi_{SNL}=\frac{1}{\sqrt{\alpha^2+\sinh^2[C\sqrt{\eta_1}]}},
\label{SNLhom}
\end{equation}
which differs from Eq.(\ref{SNLalpha}) because of the presence of one PDC term, selected by the filter. By comparing Eq.(\ref{SNLalpha}) with Eq.(\ref{SNLhom}) it is clear that the SNL in the latter case grows slowly, and the approximation $\alpha^2\gg N_{pdc}$ can hold for higher gain.
\begin{figure}[H]
	\centering
	\includegraphics[width=1\linewidth]{phasevsgainalpha}
	\caption{Optimised phase sensitivity for different gain when the seeding state consist of a coherent state in the first Schmidt mode. Comparison between direct detection and homodyne detection when the first Schmidt mode is filtered. Number of coherent photons $\alpha^2=1M$.}
	\label{alphahom}
\end{figure}
\fi
According to our analysis, this seeding case does not present any supersensitivity behaviour, and the phase sensitivity curves, very similar to Fig.\ref{salphap} and therefore not shown, lies entirely above the SNL. This result, obtained through the homodyne detection, together with the previous outcomes achieved via direct detection, suggests that a strong coherent seed in the spectral mode does not provide any improvement in the supersensitivity and therefore other seeding strategies are recommended.

\subsubsection{Coherent state in the plane-wave mode} 
In this section we consider an intense coherent seed in the plane-wave mode $\omega_p/2$. The output radiation is filtered in the same frequency. The input state in this case is given by: 
\begin{equation}
\vert\psi\rangle_{in}= \int d\omega_s\delta(\omega_s-\omega_p/2)\rvert\alpha\rangle_{\omega_s}\vert 0\rangle_{\omega_i},
\label{seedCF}
\end{equation}
where we again assume a seeding state containing one million photons. In order to prevent any time dependence in the output, the local oscillator is also defined in a plane-wave mode at the frequency $\omega_p/2$. The homodyne operator therefore takes the following expression:


\begin{equation}\begin{split}
\hat  H_d=\\
\lvert\beta_{lo}\rvert\left(e^{i\theta_a}\hat  a^{out}\left(\frac{\omega_p}{2}\right)+e^{-i\theta_a}\left[\hat  a^{out}\left(\frac{\omega_p}{2}\right)\right]^\dagger\right),
\end{split}\end{equation}
where the output plane wave operators $\hat  a^{out} $ and $\left[\hat  a^{out} \right]^\dagger$ are defined in the Appendix.

The average value of the homodyne operator gives:
\begin{equation}\begin{split}
\langle \hat H_d\rangle=2\alpha\lvert\beta_{lo}\rvert\cos\theta_a\\
\times\left(1+\sum_k\left\lvert u_k\left(\frac{\omega_p}{2},\phi\right)\right\rvert^2(\cosh\gamma_k-1)\right),
\end{split}\end{equation}
while the variance of the homodyne operator takes the form:

\begin{equation}\begin{split}
\langle \Delta^2 \hat  H_d\rangle=\\
\lvert\beta_{lo}\rvert^2 \left(1+2\sum_k\left\lvert u_k\left(\frac{\omega_p}{2},\phi\right)\right\rvert^2\sinh^2\gamma_k\right),
\end{split}\end{equation}
where the integrals in the frequency domain have already been performed.

\begin{figure}[H]
	\centering
	\includegraphics[width=1\linewidth]{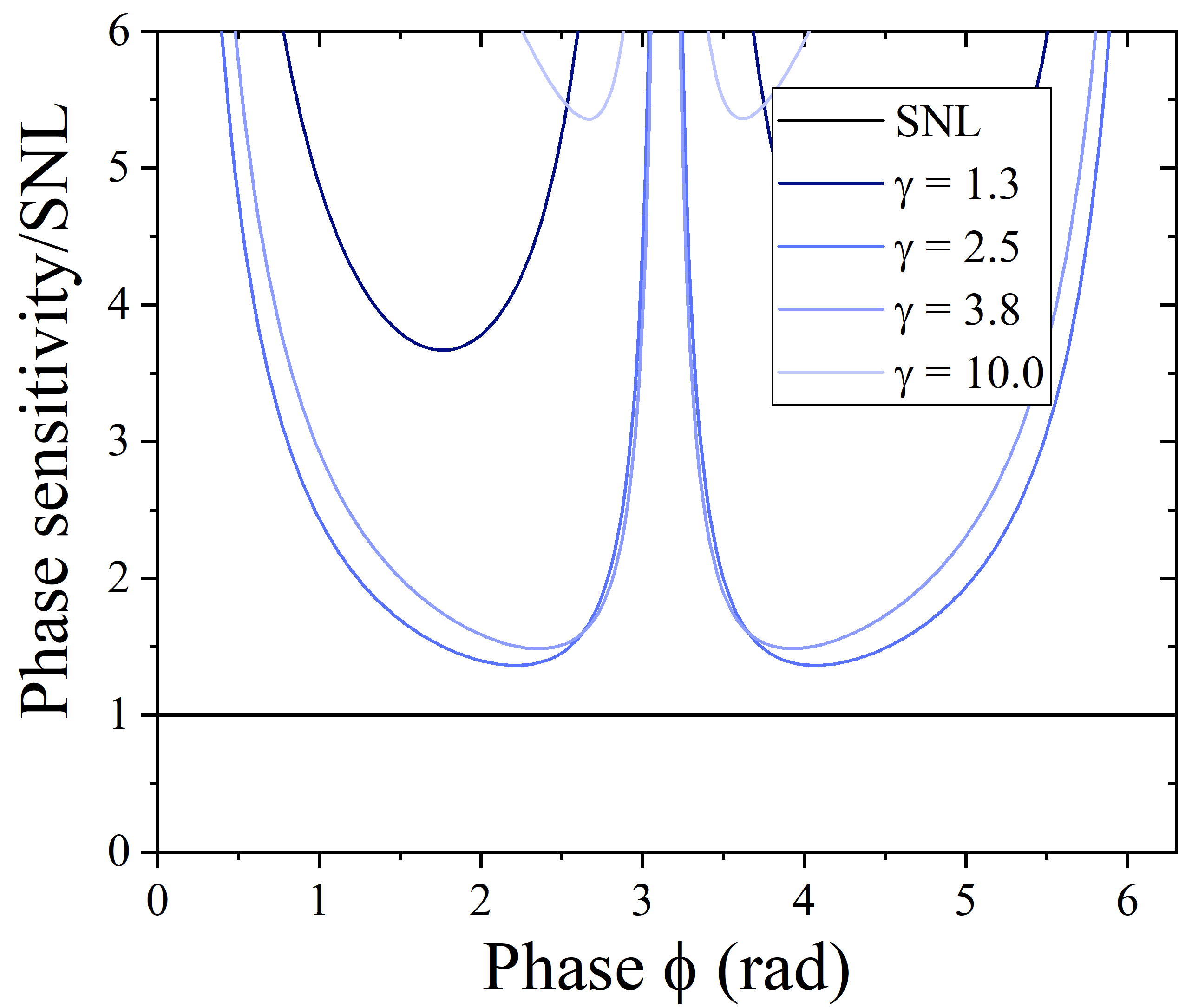}
	\caption{The normalized phase sensitivity in relation to the phase. The plane wave mode with the frequency $\omega_p/2$ is seeded by the intense coherent light having  $\lvert \alpha\rvert^2=10^6$ photons. The SNL is represented by a solid black line.}
	\label{alphaCF}
\end{figure}
\begin{figure}[H]
	\centering
	\includegraphics[width=1\linewidth]{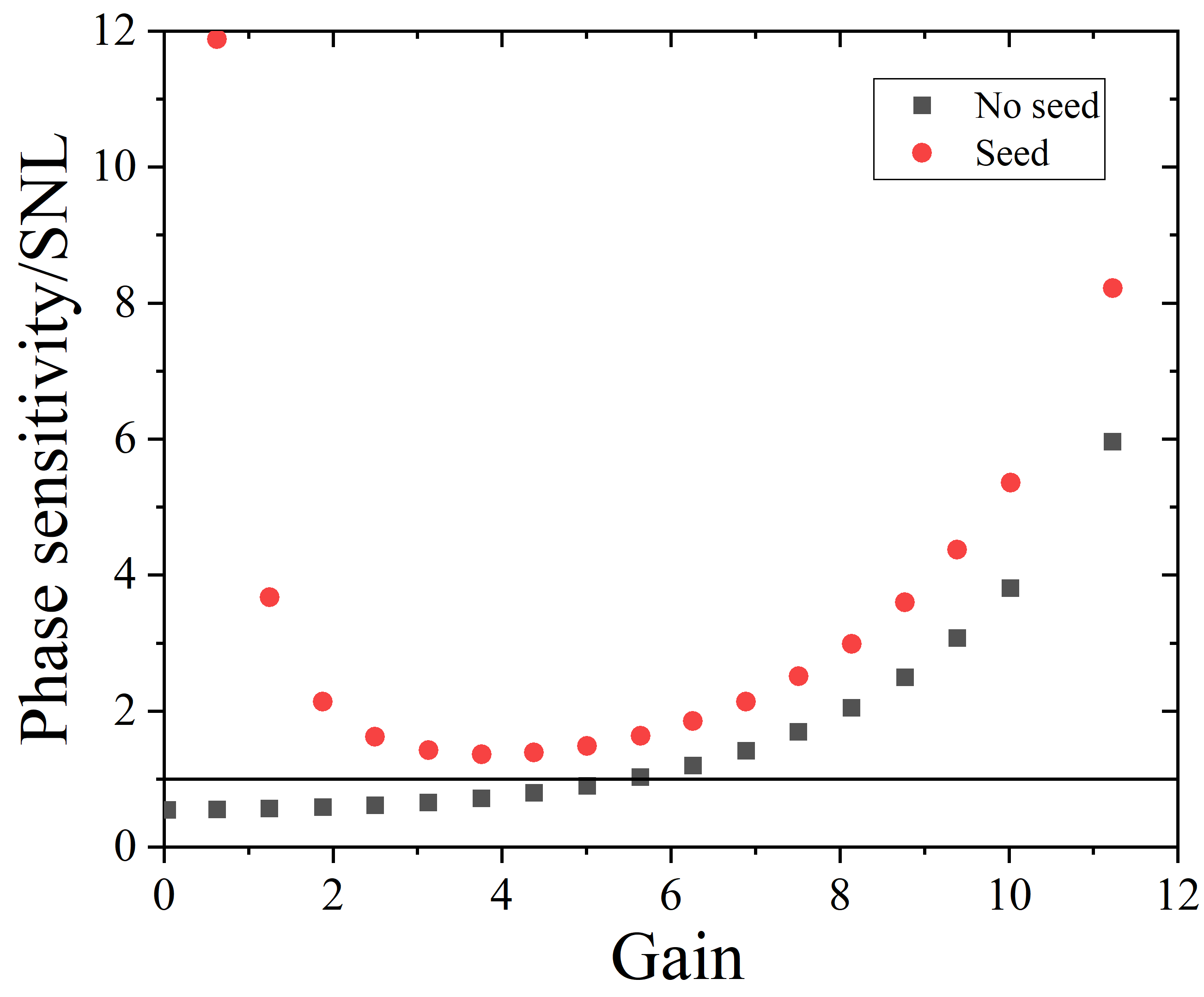}
	\caption{The minimum values of the normalized phase sensitivity presented in Fig. \ref{alphaCF}  versus gain (red circles) compared to the no-seeding case (black squares). The SNL is plotted in black. }
	\label{alphaCF2}
\end{figure}

The calculation of the shot noise limit is carried out via Eq.(\ref{snl}), taking into account the number of photons inside the interferometer after the frequency integration:
\begin{equation}\begin{split}
N_{in}=\alpha^2+\sum_k\left\lvert \bar u_k\left(\frac{\omega_p}{2}\right)\right\rvert^2\sinh^2[G_1\sqrt{\eta_k}] \\
+\alpha^2\left [\sum_k\left\lvert \bar u_k\left(\frac{\omega_p}{2}\right)\right\rvert^2(\cosh[G_1\sqrt{\eta_k}]-1) \right]^2\\
+2\alpha^2\sum_k\left\lvert \bar u_k\left(\frac{\omega_p}{2}\right)\right\rvert^2(\cosh[G_1\sqrt{\eta_k}]-1).
\end{split}
\end{equation}
According to these expressions, the phase sensitivity was calculated, normalized with respect to the SNL, optimized for $\theta_a=0$, and finally plotted in Fig.\ref{alphaCF}. The minimum values of the phase sensitivity for different gains are presented in  Fig.\ref{alphaCF2}.
This figure demonstrates that the phase sensitivity with this seeding choice also does not have any supersensitivity regions. The reason is that a strong seeding in the signal mode alone creates an imbalance between the number of signal and idler photons and destroys the signal-idler photon correlations required for the observation of the supersensitivity in the highly multimode regime.

\end{document}